\documentclass[a4paper,nobind]{ociamthesis}

\usepackage{graphicx}
\usepackage{booktabs}
\usepackage{multirow}
\usepackage{siunitx}
\usepackage{grffile}
\usepackage{subcaption}
\usepackage{capt-of}
\usepackage{bm}
\usepackage{amssymb}
\usepackage{amsmath}
\usepackage{hyperref}
\usepackage{mathtools}
\usepackage{isomath}
\usepackage{algorithm}
\usepackage[noend]{algpseudocode}

\usepackage{tabularx}
\usepackage{mwe}
\usepackage{comment}

\usepackage[style=numeric-comp, sorting=none, backend=bibtex, doi=false, isbn=false, maxbibnames=99]{biblatex}
\newcommand*{\bibtitle}{References}

\title{Improving the Security of Smartwatch Payment with Deep Learning}
\author{George Webber}
\college{St Anne's College}

\renewcommand{\submittedtext}{Submitted in partial completion of the}

\degree{MMathCompSci Mathematics and Computer Science}

\degreedate{Trinity 2023}

\DeclarePairedDelimiter{\floor}{\lfloor}{\rfloor}

\newcolumntype{Y}{>{\centering\arraybackslash}X}

\begin{document}

\setlength{\textbaselineskip}{22pt plus2pt}

\setlength{\frontmatterbaselineskip}{17pt plus1pt minus1pt}

\setlength{\baselineskip}{\textbaselineskip}

\setcounter{secnumdepth}{2}
\setcounter{tocdepth}{1}

\begin{romanpages}

\maketitle

\begin{acknowledgements}
 	I would like to express my sincere thanks to Jack Sturgess for supervising this project, and for sharing his expertise on wearable security.

I also wish to thank my friends and my family for their unwavering support throughout the highs and lows of my time at Oxford.
\end{acknowledgements}

\begin{abstract}
	Making contactless payments using a smartwatch is increasingly popular, but this payment medium lacks traditional biometric security measures such as facial or fingerprint recognition. In 2022, Sturgess et al. proposed WatchAuth, a system for authenticating smartwatch payments using the physical gesture of reaching towards a payment terminal. While effective, the system requires the user to undergo a burdensome enrolment period to achieve acceptable error levels.

In this dissertation, we explore whether applications of deep learning can reduce the number of gestures a user must provide to enrol into an authentication system for smartwatch payment. We firstly construct a deep-learned authentication system that outperforms the current state-of-the-art, including in a scenario where the target user has provided a limited number of gestures.

We then develop a regularised autoencoder model for generating synthetic user-specific gestures. We show that using these gestures in training improves classification ability for an authentication system. Through this technique we can reduce the number of gestures required to enrol a user into a WatchAuth-like system without negatively impacting its error rates.
\end{abstract}

\dominitoc 

\flushbottom

\tableofcontents

\begin{mclistof}{List of abbreviations}{3.2cm}

\item[AUROC] Area Under Receiver Operating Characteristic curve

\item[CNN] Convolutional Neural Network

\item[DTW] Dynamic Time Warping

\item[EER] Equal Error Rate

\item[FAR] False Acceptance Rate

\item[FAR@0] False Acceptance Rate when a classifier's decision threshold is optimised for zero False Rejection Rate

\item[FID] Fréchet Inception Distance

\item[FRR] False Rejection Rate

\item[GRU] Gated Recurrent Unit

\item[LSTM] Long Short-Term Memory

\item[MLP] Multi-Layer Perceptron

\item[MRR] Mean Reciprocal Rank

\item[MSE] Mean Squared Error

\item[NFC] Near-Field Communication

\item[PCA] Principal Component Analysis

\item[RF] Random Forest

\item[RNN] Recurrent Neural Network

\item[t-SNE] t-distributed Stochastic Neighbour Embedding 

\item[TSTR] Train-Synthetic Test-Real

\item[VAE] Variational Autoencoder

\item[WAE] Wasserstein Autoencoder

\end{mclistof}

\end{romanpages}

\flushbottom

\chapter{\label{ch:1-intro}Introduction} 

\minitoc

\section{Motivation}

Making contactless payments using near-field communication (NFC) is commonplace in today's society. Wearable devices such as smartwatches are an increasingly popular medium for utilising contactless payment technology --- US payments made by smartwatch are projected to increase from $ \sim $ US\$40 billion in 2020 to $ \sim $ US\$500 billion by 2024 \cite{PayingYourWrist2023}.

This rise in adoption drives a need for biometric smartwatch authentication to prevent fraudulent users making payments. Smartwatches typically lack sophisticated fingerprint sensors or cameras, but usually include an accelerometer and gyroscope for motion sensing.

\begin{figure}
	\centering
	\includegraphics[width=0.6\linewidth]{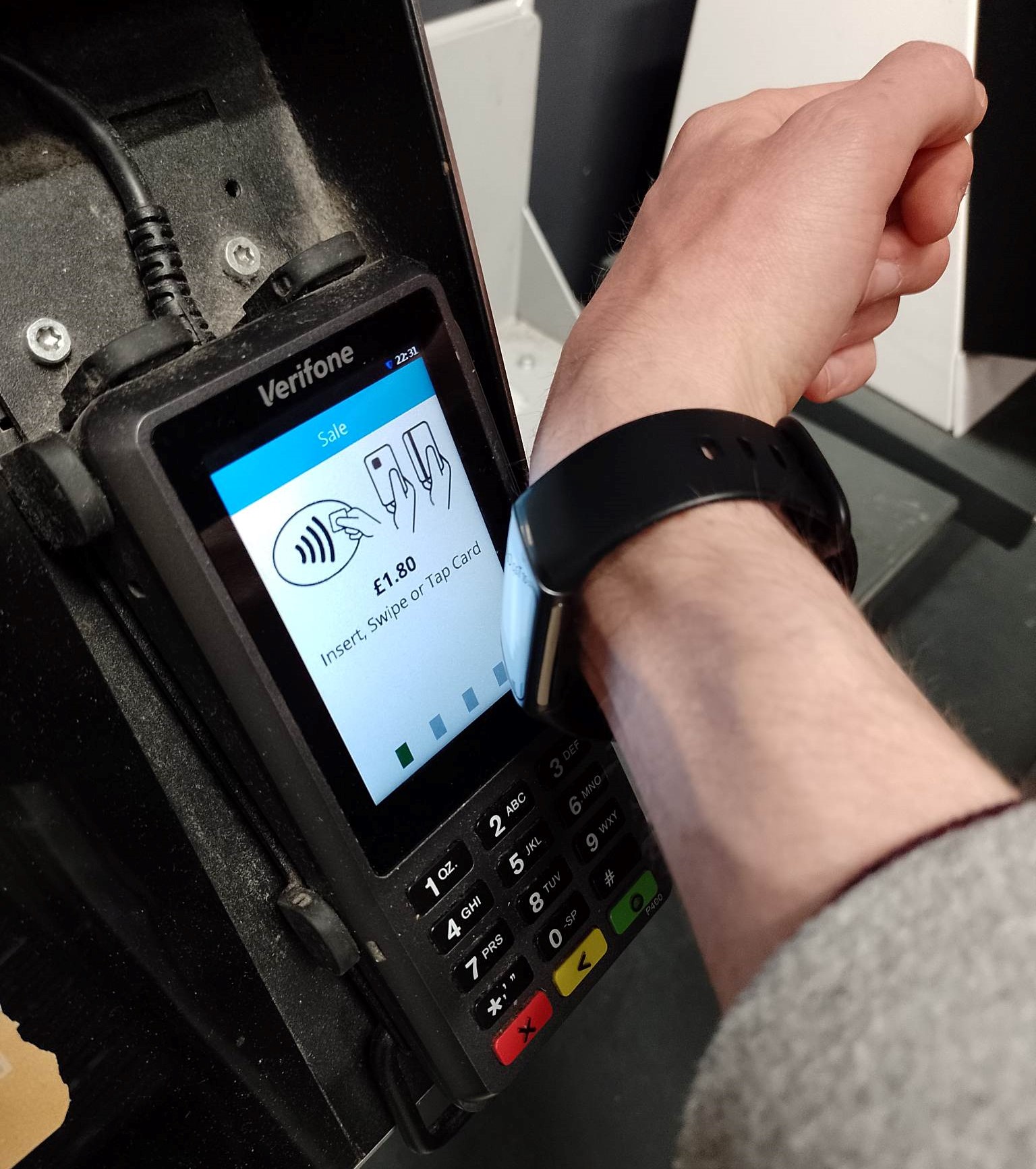}
	\caption{\small A smartwatch user making a "payment gesture".}
	\label{fig:smartwatchpaymentgesture}
\end{figure}

In 2022, Sturgess et al. introduced WatchAuth \cite{sturgessWatchAuthUserAuthentication2022}, a machine learning system for validating intent-to-pay and authenticating a user's payment based on the physical act of extending the smartwatch to a payment terminal, a motion we call a "payment gesture" (see Figure \ref{fig:smartwatchpaymentgesture}).

The attractiveness of smartwatch payments lies mostly in convenience, and so WatchAuth was conceived to improve the security of a payment without negatively affecting usability. However, similar to many biometric authentication systems, it requires a burdensome enrolment phase to become a useful security measure. In this phase, a user must provide many example gestures (on the order of 100) to create a unique template, which the system then compares to future gestures to authenticate them. Reducing the burden of this enrolment phase would encourage users to use the additional security on offer.

Deep learning is a field that has already revolutionised face and voice recognition and has the potential to improve other biometric modalities, including physical gesture recognition. In this dissertation, we investigate applications of deep learning to the task of smartwatch payment authentication, placing a particular focus on investigating ways to reduce a user's enrolment burden.

\section{Contributions}

The contributions of this work are as follows:

\begin{itemize}
	\item We survey related literature on biometric authentication and deep generative modelling for data augmentation.
	\item We reproduce the methodology and key findings of Sturgess et al.'s WatchAuth.
	\item We evaluate deep learning architectures for user authentication using the WatchAuth dataset. We find that our best architecture outperforms WatchAuth.
	\item We design and evaluate a deep-learned model for generating user-specific synthetic payment gestures. Empirical results prove we can reduce a user's enrolment burden without compromising security by augmenting the training process for a simple authentication classifier with synthetic data.
\end{itemize}

\section{Outline}

\begin{itemize}
	\item Chapter \ref{ch:2-related_work} outlines related work from the biometric security and deep learning literature, contextualising our approach.
	\item Chapter \ref{ch:3-background} gives background on topics including the WatchAuth dataset, authentication systems and generative deep learning.
	\item Chapter \ref{ch:4-auth_improvements} evaluates the effectiveness of supervised deep learning algorithms for authentication on the WatchAuth dataset.
	\item Chapter \ref{ch:5-generative_modelling} details the process of constructing a deep learning model for generating synthetic payment gestures.
	\item Chapter \ref{ch:6-evaluating_gestures} empirically investigates whether training with synthetic gestures improves a simple authentication system.
	\item Chapter \ref{ch:7-conclusion} concludes the report with an appraisal of the project and a consideration of future work.
\end{itemize}

\chapter{Related work}\label{ch:2-related_work}

\minitoc

\section{WatchAuth} \label{sec:2WA}

Sturgess et al. show that a payment gesture, performed when a user taps a smartwatch against an NFC-enabled payment terminal, is a biometric that can authenticate a user and validate their intention to pay \cite{sturgessWatchAuthUserAuthentication2022}. The system they describe for this purpose is called WatchAuth, and can be run on a smartwatch without modifying payment terminals or smartwatch hardware. In this work, we focus on the authentication use case of WatchAuth, because the intent-to-pay use case is user-agnostic and so does not require any enrolment phase.

WatchAuth's system model assumes a user is wearing a smartwatch on their wrist and uses it to make an NFC-enabled payment at a point of sale terminal in a typical setting (for example, a shop or entry barriers to a transport system). It is further assumed that the user performs a payment gesture by moving their wrist towards the terminal until it is near enough to exchange data via NFC ($ < 10$ cm), at which point the gesture ends and payment is either approved or denied. WatchAuth only requires access to the watch's inertial sensors.

To classify payment gestures, WatchAuth first windows a gesture's data to the 4 seconds before NFC contact. It then extracts manually-defined features, for example, mean, variance, skew and peak count. These features are calculated along each of the gesture data's channels (e.g. acceleration x-axis, gyroscope z-axis, acceleration Euclidean norm, etc.). A random forest classifier with 100 decision trees is then trained to classify these features, using the target user's data as the positive class and other users' data as the negative class.

\section{Biometric authentication}

\subsection{Smartwatch approaches}

There have been a number of works using inertial sensors for \textit{continuous} authentication on a smartwatch \cite{migliardiContinuousAuthenticationSmartwatch2022} \cite{acarWACAWearableAssistedContinuous2018} \cite{al-naffakhActivityBasedUserAuthentication2020} \cite{johnstonSmartwatchbasedBiometricGait2015}. For payment authentication, these approaches impose an unnecessary burden on hardware, as they run continuously instead of only when a payment is made.

Other works authenticate a user by asking them to make a simple predefined gesture (e.g. air writing or punching) \cite{yangMotionAuthMotionbasedAuthentication2015} \cite{liangUserAuthenticationWearableDevices2017} \cite{mondolUserAuthenticationUsing2017}. These systems are less convenient than WatchAuth, as they require the user to make an additional gesture to authenticate payment.

As well as inertial sensors, other smartwatch authentication schemes make use of the body's response to vibration \cite{leeUsableUserAuthentication2021}, tapping rhythms \cite{zhangSmartwatchUserAuthentication2021} or rhythmic changes in blood pressure \cite{zhaoTrueHeartContinuousAuthentication2020}. All have drawbacks in terms of convenience or battery consumption for our use case.

At present, WatchAuth is the only published approach for authenticating smartwatch payments based on a payment gesture alone, thereby offering the greatest convenience to the user.


\subsection{Deep learning approaches} \label{sec:2dl}

Most published approaches for biometric authentication using inertial sensors follow a traditional machine learning methodology similar to WatchAuth, i.e. classifying vectors of manually defined features using a random forest or support vector machine. Some recent work has seen success applying deep learning algorithms to inertial sensor data directly, including for authentication tasks. 

Both Neverova et al. and Mekruksavanich \& Jitpattanakul identify Long Short-Term Memory (LSTM) networks and their convolutional extensions as strong candidates for deep-learned authentication systems using smartphone sensor data \cite{neverovaLearningHumanIdentity2016} \cite{mekruksavanichDeepLearningApproaches2021}.

Hu et al. propose a deep learning solution for extracting optimal features from smartphone sensor data
\cite{huMultisensorBasedContinuousAuthentication2023}. Li et al. follow a similar methodology, but also incorporate a deep generative model to augment data and aid the extraction of optimal features \cite{liCNNBasedContinuousAuthentication2022}.

Smart\textit{watch} methods have seen comparatively less attention. Lu et al. introduce a deep-learned secondary authentication factor based on the user's unique motion patterns when entering a password \cite{luDeepauthInSituAuthentication2018}. Mekruksavanich \& Jitpattanakul and Yu et al. separately propose deep networks for authenticating smartwatch users using predefined custom hand gestures \cite{mekruksavanichDeepResidualNetwork2022} \cite{yuThumbUpIdentificationAuthentication2020}.

From this review, it is clear that deep learning approaches for smartwatch authentication are both realistic and relatively unexplored.



\section{Data augmentation for timeseries}

Increasing the quantity of training data available to a machine learning model improves its generalisation ability. However, in many real-world scenarios it is not possible or desirable to collect additional data directly. In this case, synthetic data can be used --- we call this process \textit{data augmentation}. 


When our data's class label is known to be invariant to certain transformations (e.g. rotations), we may apply these transformations to generate synthetic data. When class-preserving transformations are not obvious, deep generative modelling methods can be used to generate synthetic data. This is the approach taken by Um et al. for Parkinson's disease monitoring using wearable sensors \cite{umDataAugmentationWearable2017}. 

In their review of timeseries data augmentation methods, Iglesias et al. \cite{iglesiasDataAugmentationTechniques2023} identify variational autoencoders (VAEs) and generative adversarial networks (GANs) as the primary methods for deep generative data augmentation. Particular works of relevance for this dissertation are Alawneh et al. \cite{alawnehEnhancingHumanActivity2021}, who use VAEs for data augmentation in a human activity recognition context, and Goubeaud et al. \cite{goubeaudUsingVariationalAutoencoder2021}, who show that data generated from a VAE improves the performance of random forest classifiers in contexts unrelated to security. No published work was found that considers the difficult case of generative data augmentation for biometric authentication using inertial sensors.

\chapter{Preliminaries}\label{ch:3-background}

\minitoc


\section{WatchAuth dataset}

\begin{figure}[h]
	\centering
	\includegraphics[width=0.7\linewidth]{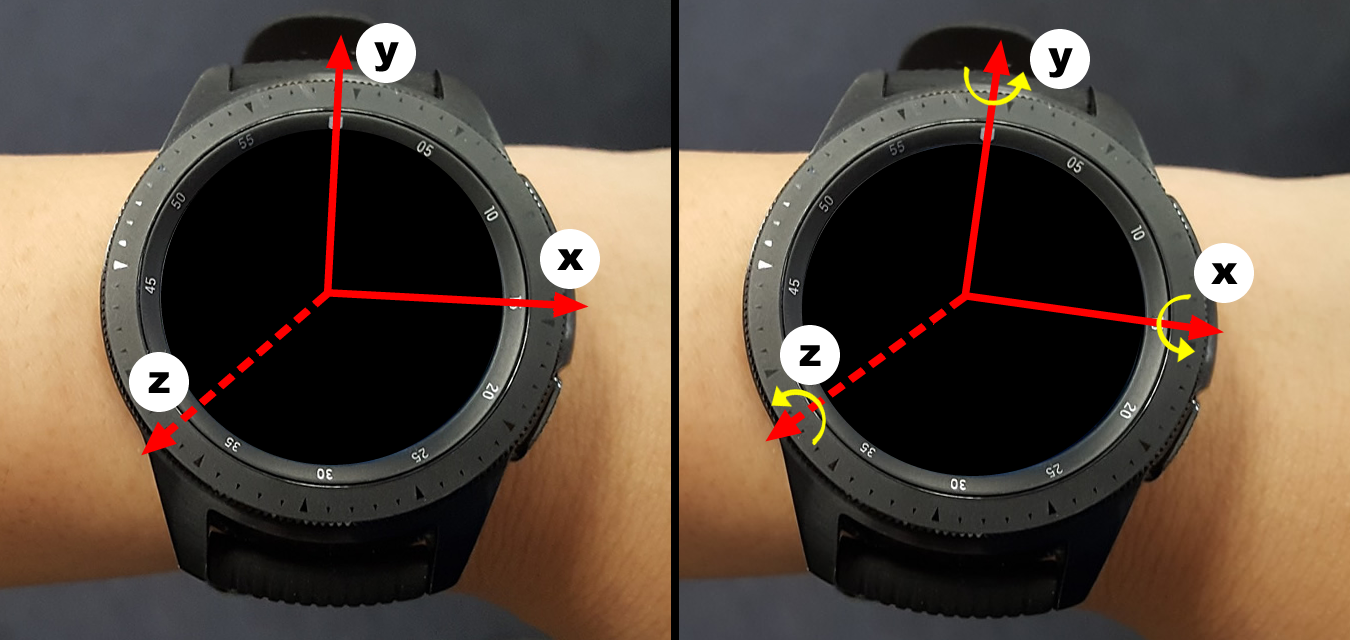}
	\caption[]%
	{{ \small The smartwatch used to collect data for the WatchAuth dataset with x, y and z axes visualised for accelerometer and gyroscope respectively (courtesy of \cite{sturgessWatchAuthUserAuthentication2022}).}}
	\label{fig:watch_axes}
\end{figure}

\begin{figure}[h]
	\centering
	\includegraphics[width=0.7\linewidth]{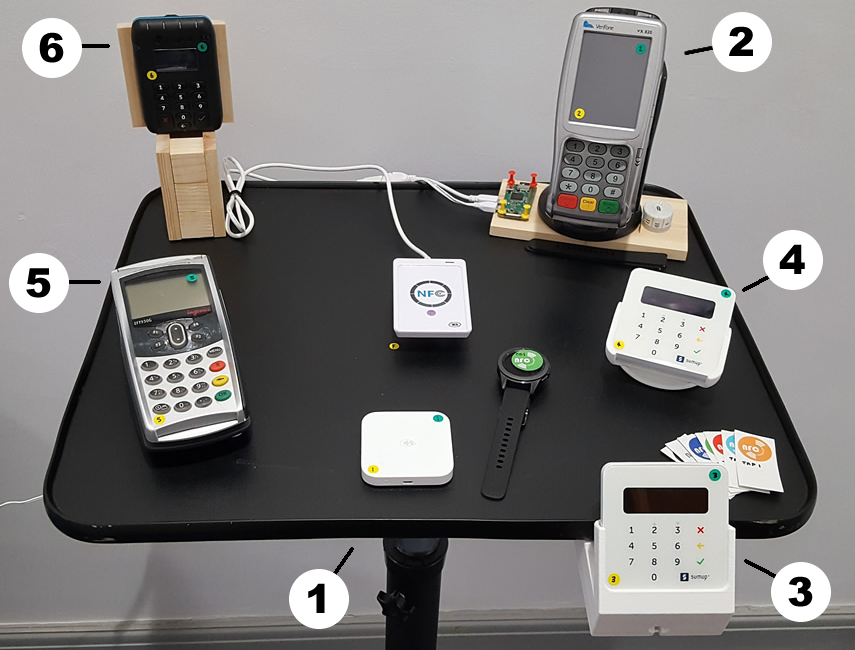}
	\caption[]%
	{{ \small The positioning of six contactless payment terminals used in WatchAuth experiments; the seventh terminal (centre) was handheld (courtesy of \cite{sturgessWatchAuthUserAuthentication2022}).}}
	\label{fig:terminal_positions}
\end{figure}

All experiments in this dissertation use the WatchAuth dataset \cite{sturgessWatchAuthDatasetExtended2023}. In March 2020, Sturgess et al. created this dataset in a physical study with 16 participants\footnote{It has been subsequently extended to 31 participants.}. Across three sessions, participants performed gestures simulating smartwatch payment at seven differently-positioned contactless payment terminals, with a watch on their wrist for data collection (see Figures \ref{fig:watch_axes} and \ref{fig:terminal_positions}). 

Data was sampled at 50 Hz from the watch's inbuilt sensors, returning vectors for acceleration (from an accelerometer), angular velocity (from a gyroscope), linear acceleration (from a linear accelerometer) and device orientation (derived from other sensors). The dataset includes timestamps showing when contact between the watch and payment terminal was initiated (via NFC).

In addition to the gesture data collected, 9 of the 16 participants collected "non-gesture" data by wearing the smartwatch in one of three public settings: in a shop, commuting on a bus or train, or while walking. This non-gesture data is useful for training classifiers for intent-to-pay recognition.

In total, 3,484 gesture and 30,771 non-gesture datapoints were collected.

Sturgess et al. show that the accelerometer and gyroscope are the most informative sensors for authentication. Furthermore, as many smartwatches only have these two sensors, this study restricts the data considered to just the acceleration and angular velocity vectors. 

The WatchAuth system itself is outlined in Section \ref{sec:2WA}.

\section{Evaluating authentication models}

Given biometric data, a biometric authentication system outputs a probability that the data belongs to a given user. If this probability is above a threshold $ T $, the system accepts the user; otherwise, it rejects them.

The system's security and usability may be quantified by its False Acceptance Rate (FAR) and False Rejection Rate (FRR) respectively. Security may be balanced against usability by modifying the threshold value $ T $.

\begin{figure}[h]
	\centering
	\includegraphics[width=0.7\linewidth]{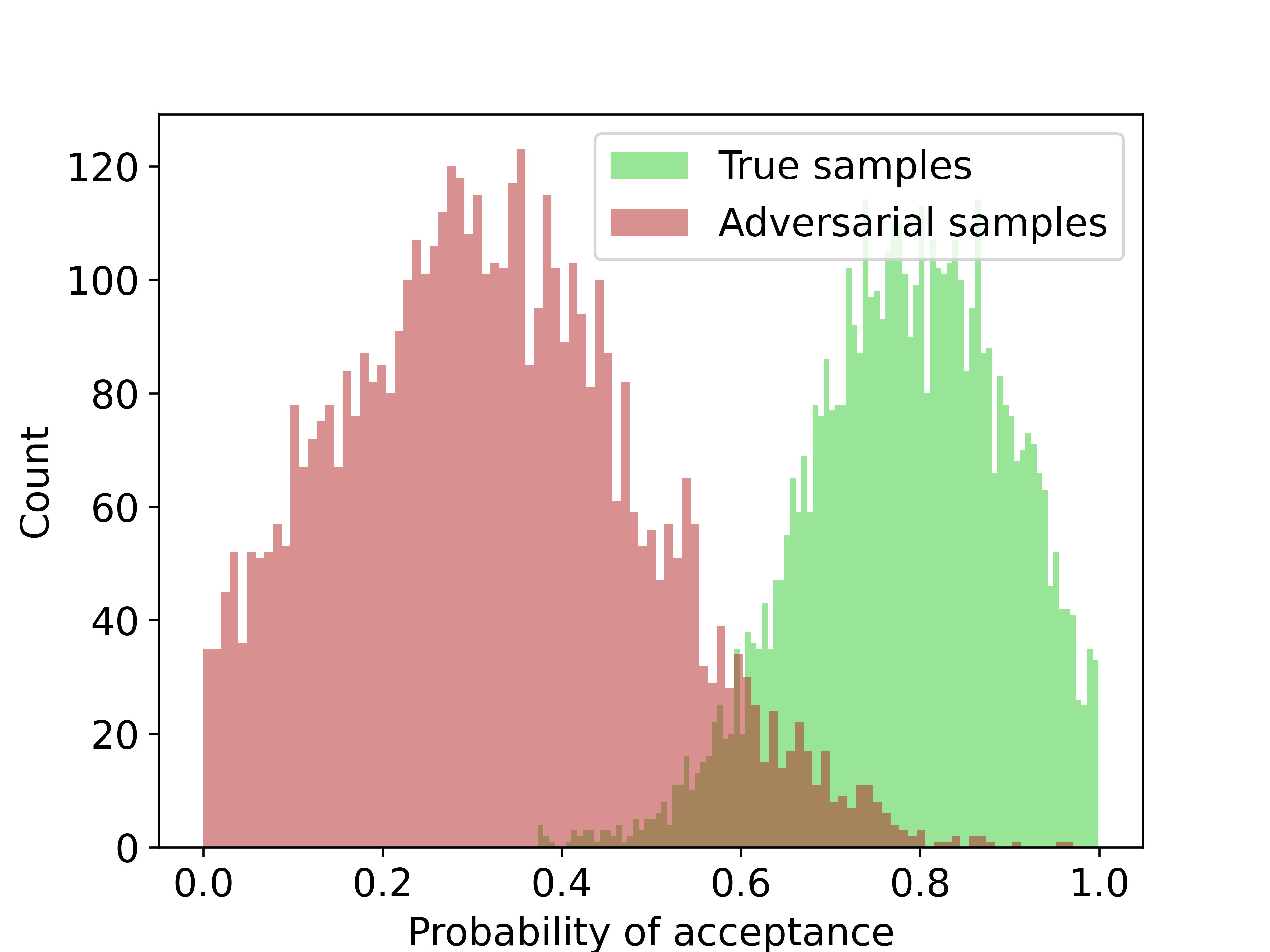}
	\caption{\small Distribution of acceptance probabilities by true and adversarial samples for an example classifier.}
	\label{fig:acceptanceprobabilities}
\end{figure}

Consider Figure \ref{fig:acceptanceprobabilities}, showing acceptance probabilities for true and adversarial data samples for an example classifier.

In this example, if we set $ T \approx 0.4 $, FRR is minimised but FAR is relatively high --- the system is maximally usable but provides little security. At $ T \approx 0.9 $, FAR is minimised but FRR is high --- the system is maximally secure but legitimate users find it difficult to gain access.

\begin{figure}[h]
	\centering
	\includegraphics[width=0.7\linewidth]{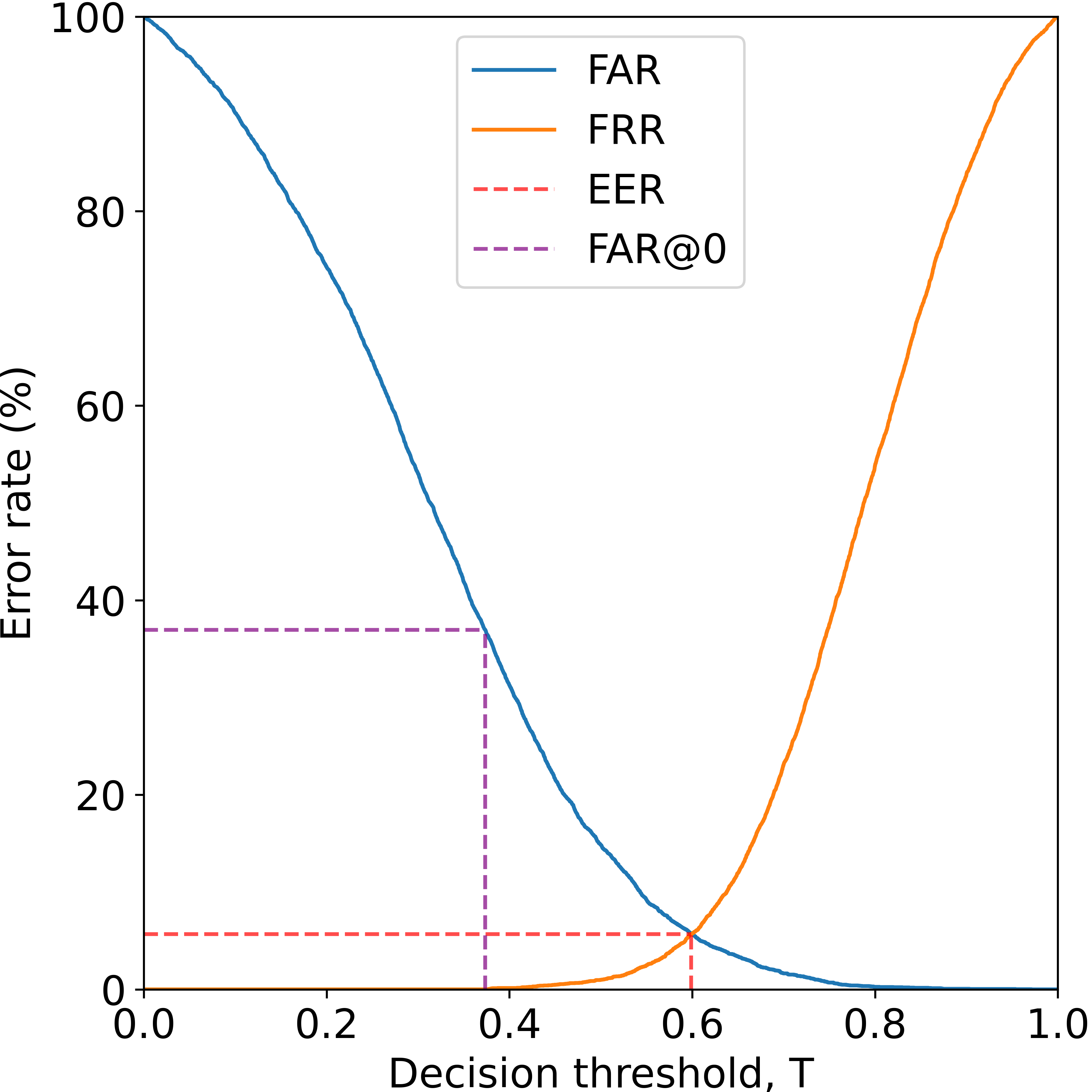}
	\caption[]%
	{{ \small FAR and FRR plotted against decision threshold $ T $ for the same example classifier as Figure \ref{fig:acceptanceprobabilities}. Note the graphical representation of the EER and FAR@0. }}
	\label{fig:eer_diagram}
\end{figure}

When we plot FAR and FRR against $ T $, we see that there exists a value of $ T $ such that FAR = FRR. The error at this point is known as the Equal Error Rate (EER). This is shown in Figure \ref{fig:eer_diagram}.

\begin{figure}[h]
	\centering
	\includegraphics[width=0.6\linewidth]{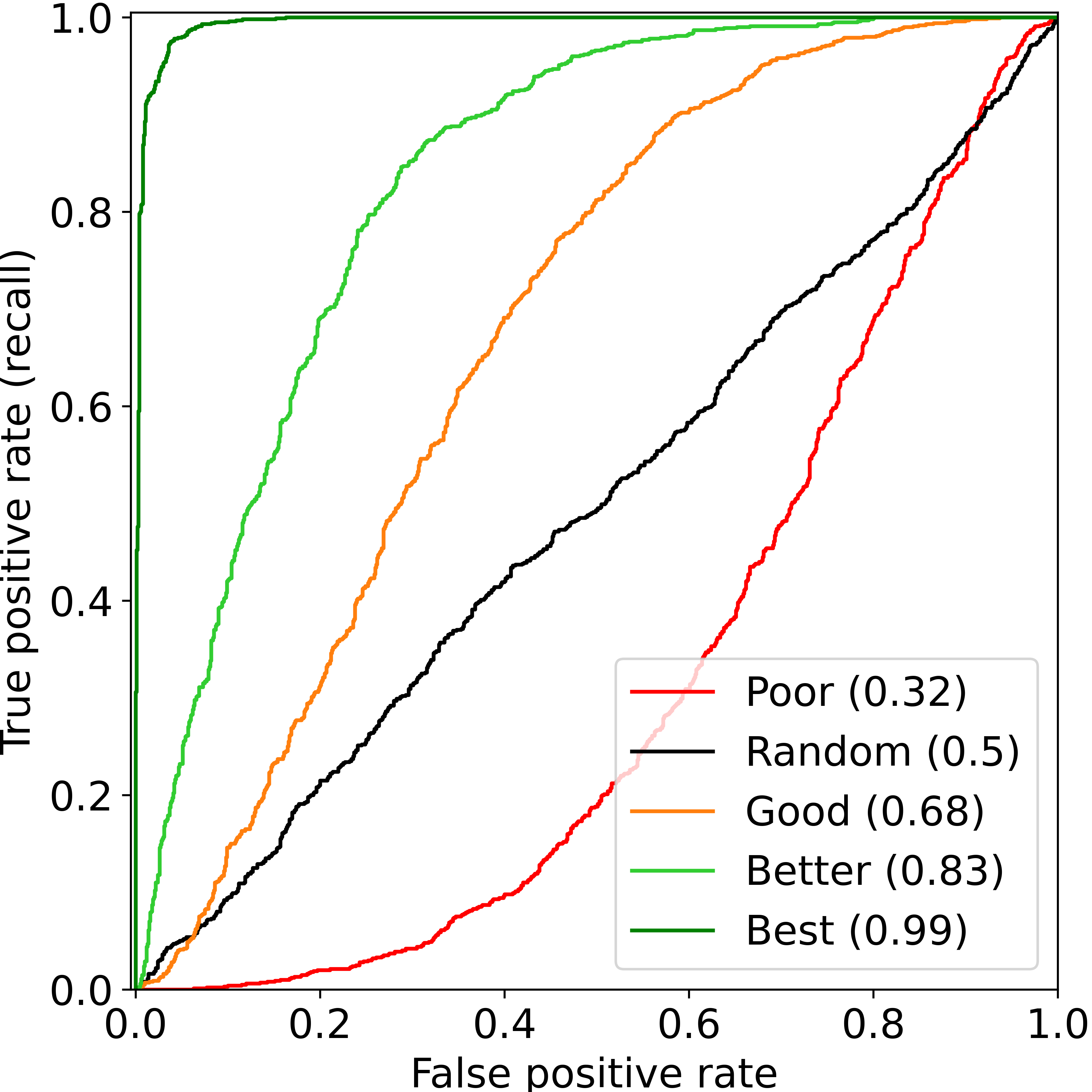}
	\caption[]%
	{{ \small ROC curves plotted for example classifiers of varying quality. AUROC values are given in brackets in the legend.}}
	\label{fig:ROC}
\end{figure}

If we instead plot the "True Accept Rate" (the proportion of true samples accepted) against FAR for different values of $ T $, we obtain the Receiver Operating Characteristic curve (see Figure \ref{fig:ROC}).

The area under this curve, known as the Area Under Receiver Operating Characteristic curve (AUROC), is another common metric for authentication. An AUROC value of 0.5 indicates a random classifier, while a value of 1 indicates a perfect classifier. Both EER and AUROC are commonly used metrics to evaluate authentication systems that quantify error independently of security versus usability considerations.

In the case of WatchAuth, maximum usability is desirable. The relevant metric in this case is the FAR when $ T $ is optimised to set FRR = 0, i.e. "the false acceptance rate when we permit no false rejections". Henceforth, we abbreviate this quantity to FAR@0.


\section{Deep learning}

\subsection{Convolutional neural networks (CNNs)}

CNNs are a popular architecture for machine learning that exploit translational invariance on grid-like domains. For 1D domains, convolving input $ \mathbold{x} $ with weight vector $ \mathbold{w} $ of size $ k $ is equivalent to a calculating a weighted moving average over $ \mathbold{x} $ with a weight vector of size $ k $. Assuming odd $ k $ and zero-padded input, convolving $ \mathbold{x} $ with $ \mathbold{w} $ yields a new vector $ (\mathbold{x} \star \mathbold{w}) $ with elements

\[ \qquad (\mathbold{x} \star \mathbold{w})_i = \sum_{j = 0}^{k} w_j \cdot x_{i + j -  \floor{\frac{k}{2}}} \qquad  .\]

In deep learning, a convolution layer convolves its input with many learnable weight vectors in parallel to extract different representations of the input data. In CNNs, maximum pooling is also typically used to reduce the spatial dimension of an input vector. When applied in sequence, convolution and maximum pooling layers allow a network to extract features at multiple scales, with increasingly complex features at higher depths.

\subsection{Gated recurrent unit}

Recurrent neural networks (RNNs) are a type of neural network designed to process sequence data. A learnable weight matrix is iteratively applied to each item in a sequence to update an internal state. When applied to long sequences, standard RNNs suffer from exploding/vanishing gradient issues when backpropagating, limiting their ability to model long-range dependencies.

Gated recurrent units (GRUs) alleviate this issue through \textit{update} and \textit{reset} gating operations. These operations decide what information from the previous state to carry forward to the new state and what to forget, thereby improving information propagation through the network \cite{choPropertiesNeuralMachine2014}.

\section{Generative deep learning}\label{sec:3generativemodels}

\subsection{Autoencoders} 

One common family of generative deep learning models are autoencoder models. Under this semi-supervised framework, we have two neural networks: an encoder $ E $ and decoder $ D $. Given input $ x $, we firstly apply the encoder $ E $ to generate an embedding in a latent space (typically a Euclidean space of low dimension). We then apply the decoder $ D $ to $ E(x)$ to reconstruct a signal in the original domain. For a given loss function $ \ell$, the objective of the network is to minimise the reconstruction loss $ \ell(x, (D \circ E) (x)) $.

An autoencoder's latent space must be sufficiently regular to be useful for generating synthetic data. In particular, we require it to be sufficiently continuous so that close together points decode to semantically similar representations \cite{roccaUnderstandingVariationalAutoencoders2021}, and to be complete, i.e. without "holes".

Unregularised autoencoders only minimise reconstruction loss, which usually results in a latent space that does not encode useful semantic information and is full of holes where the decoder has never been trained \cite{tolstikhinWassersteinAutoEncoders2018}.

\subsection{Variational autoencoders} \label{sec:3vaes}

\begin{figure}
	\centering
	\includegraphics[width=0.7\linewidth]{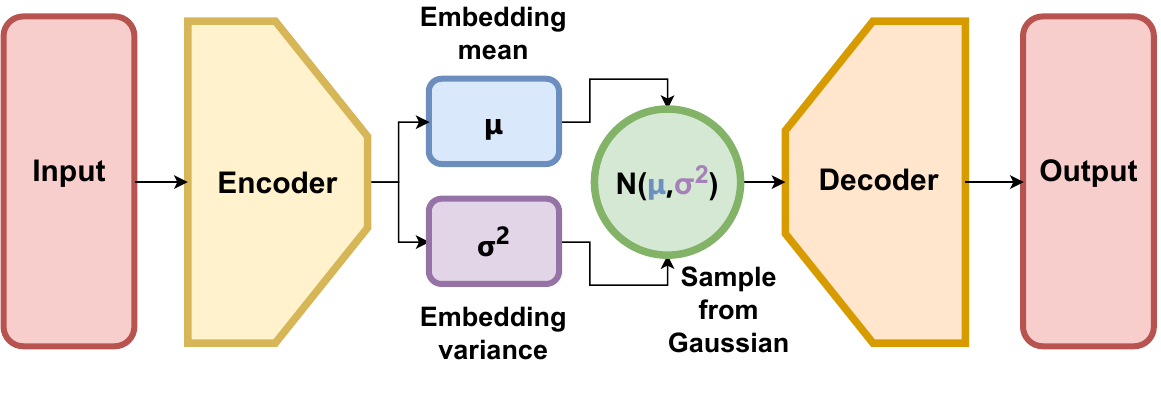}
	\caption{\small The process of embedding and reconstructing a point using a VAE.}
	\label{fig:vaediagram}
\end{figure}

First proposed by Kingma \& Welling, variational autoencoders (VAEs) are the most well-known version of a regularised autoencoder \cite{kingmaAutoEncodingVariationalBayes2014a}. Instead of outputting a single vector in the latent space, the encoder is modified to output a latent mean and variance representing the parameters of a Gaussian distribution. The decoder then samples from this Gaussian and decodes the sampled point\footnote{This is a simplification --- in fact, any distribution may be used but a Gaussian is the natural choice due to its relative tractability.}. Figure \ref{fig:vaediagram} shows this process.

To comply with our latent space requirements of continuity and completeness, we want our latent variances to be sufficiently large and our latent means sufficiently close such that the latent distributions of different points overlap. We achieve this using the Kullback-Leibler (KL) divergence to penalise the distance between each latent distribution and a standard Gaussian. This gives rise to the following KL loss used as a regularisation term in the VAE loss function\footnote{In practice, our encoder outputs the \textit{logarithm} of the latent variance, resulting in a change of variables in this formula.}:

\begin{align*}
	\ell_{KL}(\mu, \sigma^2) &= D_{KL}(\mathcal{N}(0,1) \space || \space \mathcal{N}(\mu, \sigma^2)) \\
	\qquad &= -\frac{1}{2} (1 + log(\sigma^2) - \mu^2 - \sigma^2) \qquad .
\end{align*}

VAEs have a strong theoretical basis and may be derived from a variational inference perspective.

The full loss function used in a VAE is the sum of the reconstruction and KL losses.

\subsection{Wasserstein autoencoders}\label{sec:3waes}

\begin{figure}
	\centering
	\includegraphics[width=0.7\linewidth]{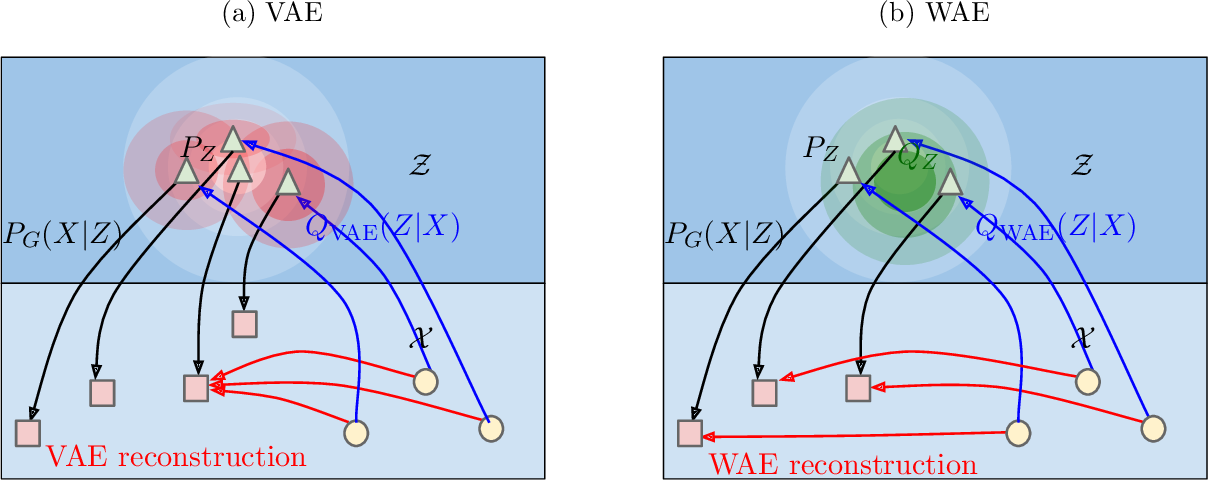}
	\caption[]{\small A visual representation of the difference between VAE and WAE regularisation (reproduced from Tolstikhin et al.). In a VAE, the divergence between each point's distribution and a Gaussian is penalised; in a WAE, the distance between the distribution of all points and a Gaussian is penalised. We also see that in the VAE case, nondeterminism leads different inputs to map to the same output, and this is avoided in the WAE case. For further details, see the original paper \cite{tolstikhinWassersteinAutoEncoders2018}.}
	\label{fig:waevsvae}
\end{figure}

Wasserstein autoencoders (WAEs), proposed by Tolstikhin et al. in 2019, are an alternative method for regularising an autoencoder (see Figure \ref{fig:waevsvae}) \cite{tolstikhinWassersteinAutoEncoders2018}.

Instead of modelling points as distributions in the latent space and regularising these individual distributions, WAEs use a regularisation term that encourages the distribution of all latent space embeddings to match a Gaussian\footnote{As with VAEs, this is a simplification. Any well-defined distribution may be used without difficulty.}. Given $ n $ training points $ \{x_1, ..., x_n\} $, and $ n $ points $ \{ z_1, ..., z_n \} $ sampled independently from a standard Gaussian, we calculate the following regularisation loss:

\begin{align*}
	\ell_{WAE\_regularisation} = \frac{1}{n(n-1)} & \sum_{i \ne j} (E(x_i) - E(x_j))^2 + \frac{1}{n(n-1)} \sum_{i \ne j} (z_i - z_j)^2 \\
	&- \frac{2}{n^2} \sum_{i,j} (E(x_i) - z_j)^2  \qquad .
\end{align*}

Tolstikhin et al. claim that WAEs retain the favourable qualities of VAEs including stable training and nice latent manifold structure while achieving higher quality reconstructions (which they demonstrate empirically for image data).

\section{Dynamic time warping}\label{sec:3DTW}

\subsection{DTW algorithm}

\begin{figure*}[t] 
	\centering
	\begin{subfigure}[b]{0.475\textwidth}
		\centering
		\includegraphics[width=\textwidth]{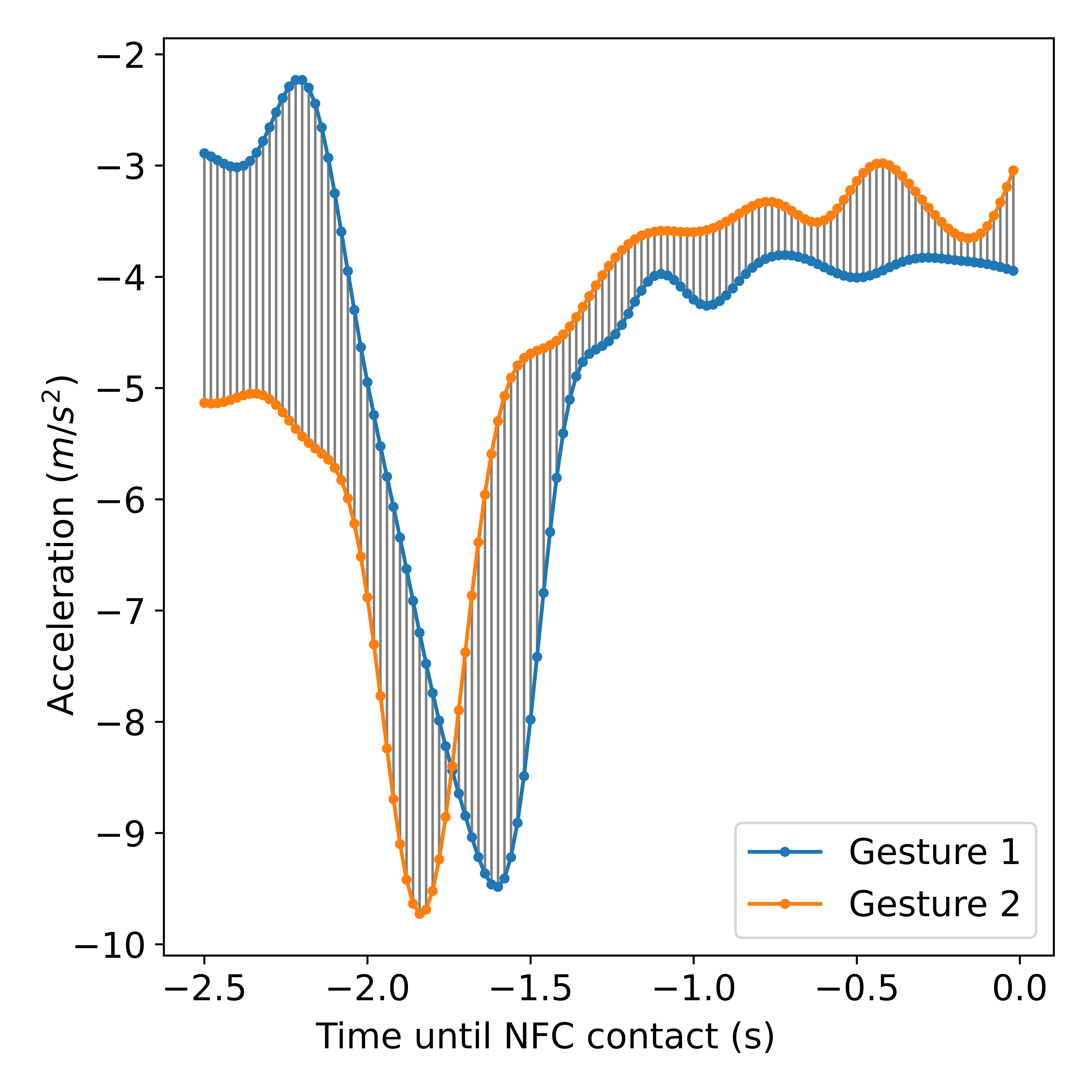}
		\caption[]%
		{{\small Euclidean distance}}
	\end{subfigure}
	\hfill
	\begin{subfigure}[b]{0.475\textwidth}
		\centering 
		\includegraphics[width=\textwidth]{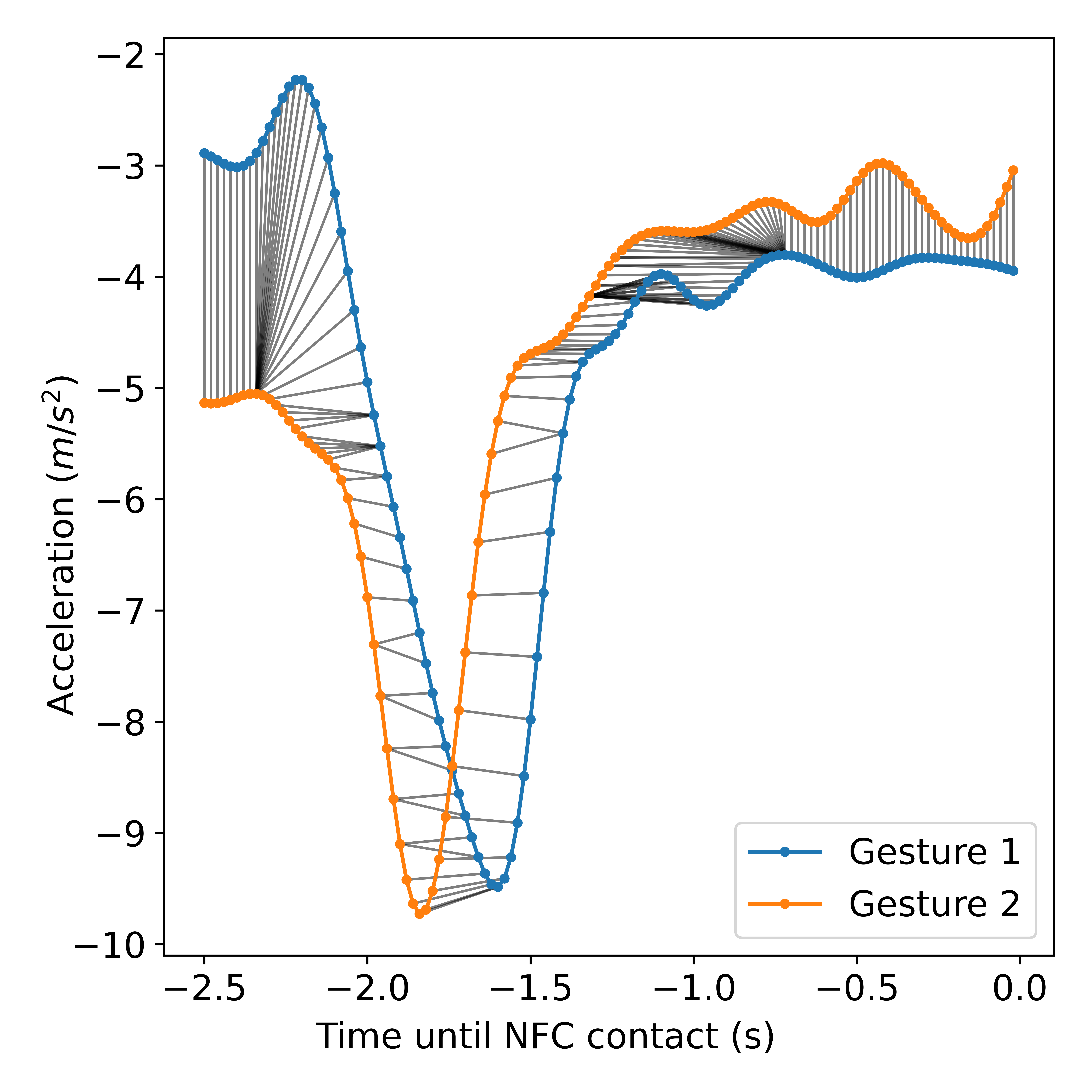}
		\caption[]%
		{{\small DTW}}
	\end{subfigure}
	\hfill
	\caption[]%
	{{\small The dissimilarity of two WatchAuth gestures, measured with (a) Euclidean distance and (b) DTW. Matching between points is shown in grey.}}
	\label{fig:dtwexplained}
\end{figure*}

Dynamic time warping (DTW) is an algorithm for quantifying the dissimilarity between two timeseries that are not perfectly aligned. The DTW distance is computed as the minimum distance between the timeseries across all temporal alignments. A given temporal alignment may be described by a matching between two sequences (see Figure \ref{fig:dtwexplained}).
Given sequences $ \mathbold{x} = (x_n)_{n=1}^M $ and $ \mathbold{y} = (y_n)_{n=1}^N $, a valid matching must obey the following rules:

\begin{enumerate}
	\item The first and last points of the sequences must be matched to each other (but may also match other points).
	\item Every index from both sequences must be matched with at least one index from the other sequence.
	\item Indices from the first sequence must be matched in a monotonically increasing manner to indices from the second sequence, and vice versa.
\end{enumerate}

The cost of a matching is the sum of distances between matched points. An optimal matching may be computed via dynamic programming in time and space complexity equal to the product of the lengths of the sequences, see Algorithm \ref{alg:DTW}.

\begin{algorithm} 
	\caption{Dynamic Time Warping}\label{alg:DTW}
	\begin{algorithmic} 
		\Function{DTW}{X: Array [0..m], Y: Array [0..n]}
		\State $D \gets $ Array [0..m, 0..n] \Comment{Initialise distance matrix.}
		\For{ $ i \gets 1 $ to $ m $ }
			\For{ $ j \gets 1 $ to $ n $ }
				\State $ D[i, j] \gets \infty$ 
			\EndFor
		\EndFor
		\For{ $ i \gets 1 $ to $ m $ }
			\For{ $ j \gets 1 $ to $ n $ } \Comment{ $ d $ is any distance function.}
			\State $ D[i,j] \gets d(X[i], Y[j]) + \min\{ D[i-1, j],  D[i, j-1],  D[i-1, j-1]\} $ 
			\EndFor
		\EndFor \\
		\Return $ D[m,n]$
		\EndFunction
	\end{algorithmic}
\end{algorithm}

\subsection{DTW variants}

\subsubsection{Sakoe-Chiba bandwidth}

Without further constraints, DTW exhibits degenerate behaviour such as matching one timeseries with just the endpoints of the other. This behaviour may be eliminated by imposing a bandwidth constraint on DTW, for example the Sakoe-Chiba bandwidth \cite{sakoeDynamicProgrammingAlgorithm1990}. This technique limits the distance between indices that may be matched to a fixed value $ w $, and may be trivially incorporated into Algorithm \ref{alg:DTW} by changing the range of the iteration over the recursion relation.

\subsubsection{Lower bounds on DTW} \label{sec:3LB_Keogh}

DTW with Sakoe-Chiba bandwidth $ w $ on sequences of equal length $ l $ requires space and time complexity $ O(w \cdot l)$. This high complexity motivated research into lower bounds on DTW that have complexity $ O(l) $, one of which is Keogh's lower bound\footnote{
More sophisticated lower bounds are an active topic of research; see Webb \& Petitjean \cite{webbTightLowerBounds2021}.}. 

Keogh's lower bound makes use of derived series, termed \textit{envelopes}, around a timeseries. For a sequence $ \mathbold{x} = (x_i)_{i=1}^n $, its upper and lower envelopes $ U_x $ and $ L_x $ are the maximum and minimum values of $ \mathbold{x} $ within a moving window of width $ w $:

\[ \qquad U_i^{\mathbold{x}} = \max_{\max\{1, i-w\} \le j \le \min \{ l, i + w \}} \{x_j\} \qquad \]
\[ \qquad L_i^{\mathbold{x}} = \min_{\max\{1, i-w\} \le j \le \min \{ l, i + w \}} \{x_j\}  \qquad . \]

\begin{figure}
	\centering
	\includegraphics[width=0.7\linewidth]{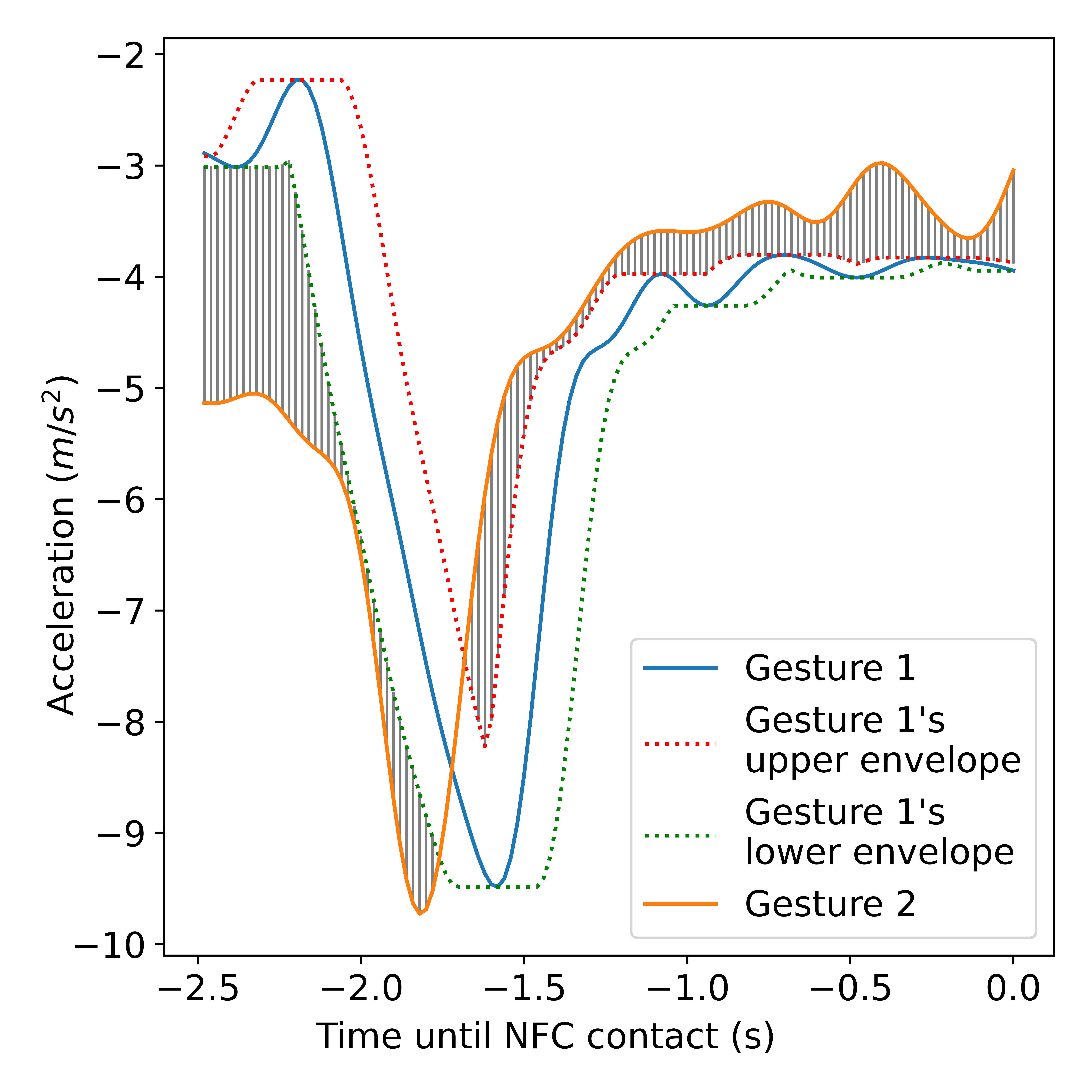}
	\caption{\small Measuring dissimilarity of two WatchAuth gestures with Keogh's lower bound.}
	\label{fig:lbkeogh}
\end{figure}

Keogh's lower bound is the sum of distances at points where sequence $ \mathbold{y} = (y_i)_{i=1}^n $ is not contained within $ \mathbold{x} $'s envelopes (see Figure \ref{fig:lbkeogh}): 
\[
\qquad d_{KLB}^{(w)}(\mathbold{x}, \mathbold{y}) = \sum_{i=1}^n \begin{cases}
	\ell(y_i, U_i^{\mathbold{x}}) \qquad	\text{if $y_i > U_i^{\mathbold{x}}$},\\
	\ell(y_i, L_i^{\mathbold{x}}) \qquad	\text{if $y_i < L_i^{\mathbold{x}}$},\\
	0 \qquad \qquad \quad \text{otherwise}
\end{cases}
\qquad .
\]

\chapter{Supervised deep learning for authentication}\label{ch:4-auth_improvements}

\minitoc
In this chapter, we investigate whether supervised deep learning classification models can outperform the existing WatchAuth system for user authentication.

\section{Methodology}

\subsection{Threat model}

We follow the threat model laid out by Sturgess et al. \cite{sturgessWatchAuthUserAuthentication2022}.

We assume that an adversary has possession of a legitimate user's unlocked smartwatch, and that the adversary attempts to make a contactless payment using it. This is a zero-effort attack, as the adversary is attempting to fool the authentication system by submitting their own biometric signature.

\subsection{Preprocessing data}

The WatchAuth dataset is provided as per-user files of partially-processed gesture data. Each row is associated with a gesture ID, timestamp and one of the watch's four sensors with sensor readings (e.g. x-value, y-value etc...).

The standard input format for timeseries prediction models is a real vector of dimension $N_{timesteps} \times N_{channels} $. We therefore pre-process each row to be indexed by gesture ID and timestep rather than by gesture ID, timestep and sensor.\footnote{Timestamps on the dataset do not match perfectly between sensors, making this a non-trivial programming task.} The data provided per gesture covers a timespan from 4 seconds before NFC contact to 2 seconds after. For real-time authentication, time after the payment is accepted is irrelevant, so we slice all gestures to only cover the 4 seconds before payment is made. Non-gesture data provided in the WatchAuth dataset is also partitioned into 4 second windows.

Restricting to the sensors of interest, we represent each gesture by a separate $200 \times 6$ array (3 channels each for accelerometer and gyroscope).
Each gesture array is then passed through a Butterworth low-pass filter to remove noise. We also store descriptive information for each gesture, including user ID and payment terminal position.

\subsection{Experiments}

The task of user authentication is to classify timeseries by whether or not they represent a payment gesture for the user in question.

The choice of experimental setup can strongly influence reported results (see Eberz et al. \cite{eberzEvaluatingBehavioralBiometrics2017}). To align with best practice, we preserved the temporal order of samples when constructing train/test splits. This is because authentication systems have distinct enrolment and authentication phases that should be preserved when modelling. The first two-thirds of all users' gestures was used for training and validation, with the final third used for testing. Deep learning models used a random train/validation split of 80\%/20\%, while WatchAuth used a train/validation split of 100\%/0\%.

For the timeseries data used by deep learning models, channel-wise means and variances were initialised using the training data and used to normalise all data prior to classification.

We investigated a range of model architectures. For each user, a model was trained to classify gestures by whether they were made by the target user, with its performance evaluated on the test dataset. Performance metrics were calculated as an average over all 16 users and 5 repetitions using random weight initialisations and random validation splits stratified by class label.

We investigated two settings: \textit{full}, with all training gestures available for training and validation; and \textit{limited}, where each model was limited to using just 10\% of a target user's training and validation gestures.

\subsubsection{Original WatchAuth approach}

For comparison purposes, we reimplemented the methodology of WatchAuth. We firstly extracted manually defined features, and then trained a random forest model with 100 decision trees to classify feature vectors. For a fairer comparison with respect to number of parameters, we also trained a 1000 tree version of the random forest model.

As these models are not identical to WatchAuth due to the exclusion of some sensors in the dataset, we refer to them as \textit{RF100} and \textit{RF1000} (for "random forest with 100 and 1000 decision trees" respectively).

\subsubsection{Implementation details}

All deep learning models were trained with the Adam optimiser (with learning rate $1e-4$) and weighted binary cross-entropy as their loss. Due to the severe class imbalance (approximately 1:15 weighted against the positive class), all deep learning models struggled to make meaningful predictions without any weight rebalancing. Empirically, weighting the training data 4:1 in favour of the positive class showed the strongest classification performance on validation data.


Hyperparameters were tuned manually to minimising average validation loss. During training, validation loss was used for model selection (using early stopping with a patience of 150 epochs). 

\subsection{Model choice}

Five deep learning models of increasing complexity were explored. For meaningful comparison, all models have $ \sim 70 $k trainable parameters. The models differ by how they embed their timeseries input into a feature representation space; all models then apply the same standard dense layers for classification and output through a single sigmoid-activated neuron. The output of each model may be interpreted as the probability that the input data is a true payment gesture for the authentication user.

In Appendix \ref{app:1-model diagrams}, we visualise each model architecture.

The most basic model, \textit{MLP}, generates its feature embeddings via stacked dense layers interleaved with ReLU non-linearity.

The \textit{ConvNet} model is more involved, stacking convolutional and maximum pooling layers sequentially. These layers obtain a feature representation with reduced spatial dimension and increased number of feature channels; this representation is then flattened before classification. This model is designed to be tolerant to shifts in the time dimension.

Theoretically, a recurrent model can more effectively exploit the sequence structure of an input timeseries than a convolutional model, by modelling intra-sequence dependencies. We designed the \textit{GRU} model to investigate whether a recurrent model performs well for authentication by stacking GRU layers\footnote{The GRU recurrent model was found to greatly outperform the LSTM recurrent model, which could not be trained effectively.}.

In addition to these simple models, we investigated combining convolutional layers and GRU layers in sequence (as is common in the literature, see Section \ref{sec:2dl}). Intuitively, this architecture can first extract translation-invariant features using convolutions, and then exploit intra-sequence dependencies with the GRU. This approach also mitigates training difficulties when using GRUs for long sequences as the convolutional layers reduce the spatial dimension of the data.

Results for two hybrid models are presented. The simpler model, \textit{SimpleMix}, applies five convolutional layers to reduce the spatial dimension of the data to seven timesteps.

The more complex model, \textit{ComplexMix}, was inspired by Chrononet \cite{royChronoNetDeepRecurrent2019a}, and employs varying kernel sizes to extract features at multiple scales. These features are then concatenated and passed through 1x1 kernels to reduce the overall number of parameters. Four layers of this architecture are employed to reduce the spatial dimension of the data to thirteen timesteps. Both \textit{SimpleMix} and \textit{ComplexMix} then pass data through three stacked GRUs before the classification MLP.

\section{Results}

\subsection{Full data}

\begin{table*}[h]
	\caption[]{\small \textit{Full} experiment results for assorted models trained using the full training dataset.} 
	\label{tbl:DL_results_summary}
	\vskip 0.15in
	\begin{center}
		\begin{sc}
			
\begin{tabularx}{\textwidth}{l|YYY}
	\toprule
	Model       & AUROC & EER     & FAR@0\\
	\midrule
	RF100       & 0.951  & 0.097  & 0.52 \\
	RF1000      & 0.956  & 0.093  & 0.37 \\
	\midrule
	MLP         & 0.932  & 0.114   & 0.49 \\
	ConvNet     & 0.948  & 0.097  & 0.45  \\
	GRU         & 0.910  & 0.137   & 0.64  \\
    SimpleMix   & 0.942  & 0.097  & 0.44   \\
	ComplexMix  & 0.956  & 0.079  & 0.39   \\
	\bottomrule
\end{tabularx}

		\end{sc}
		\vskip -0.4in
	\end{center}
\end{table*}

Table \ref{tbl:DL_results_summary} shows the mean value of authentication metrics for each model evaluated on the test dataset.

It may be observed that \textit{ComplexMix} was the only deep learning model to outperform the original WatchAuth method (\textit{RF100}) across all metrics. 

Adding more trees to the random forest (\textit{RF1000} vs \textit{RF100}) improved performance on outliers as shown by the improvement in FAR@0, but had limited effect on EER or AUROC.

It is also noteworthy that the \textit{GRU} model has worse performance than the \textit{MLP} baseline --- a likely explanation for this is that the input timeseries are too long for GRU layers to effectively exploit intra-sequence dependencies (due to the vanishing gradient problem).

\subsection{Limited data} \label{sec:4limiteddata}

\begin{figure}[h]
	\centering
	\includegraphics[width=0.5\linewidth]{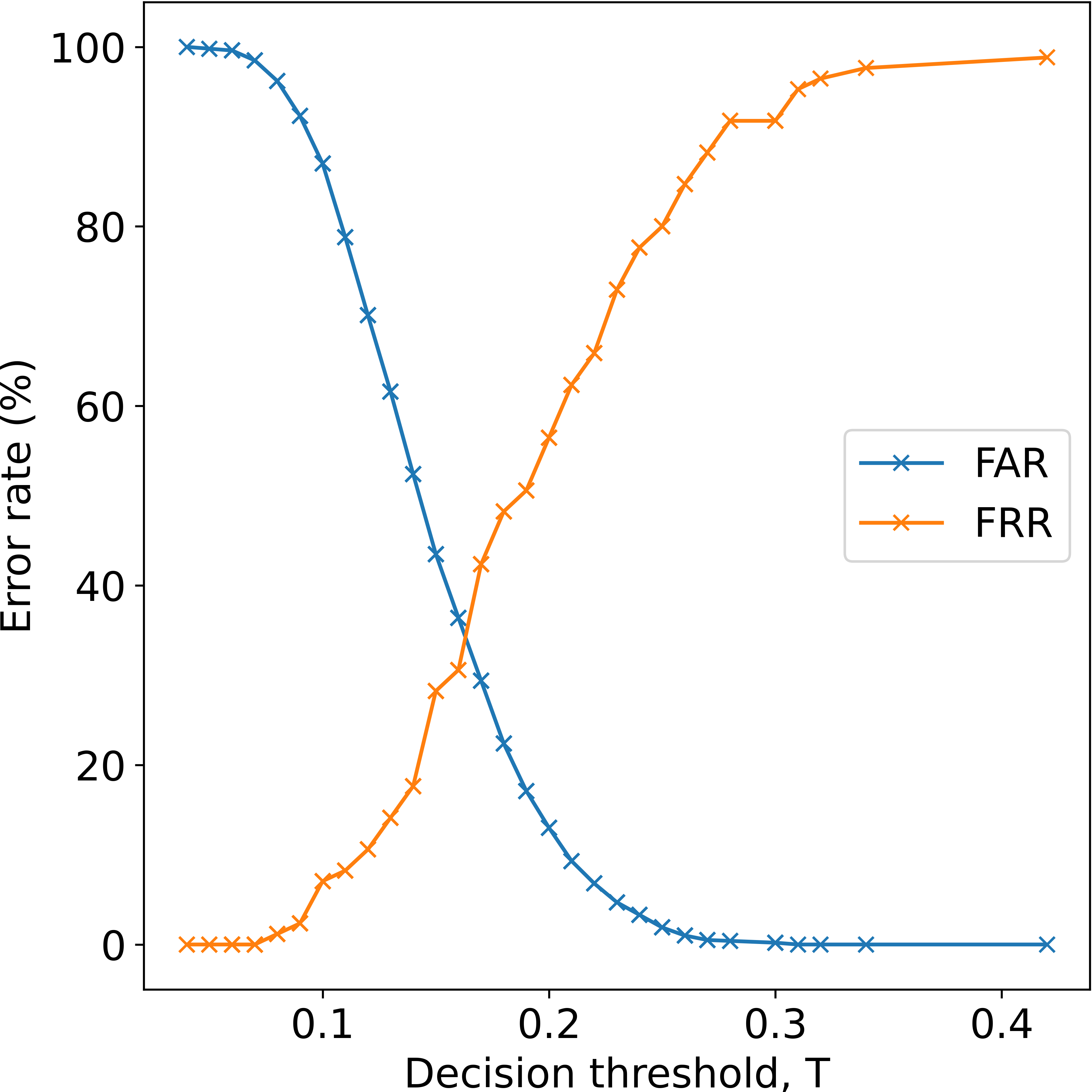}
	\caption{\small A plot of FAR and FRR against decision threshold $ T $ from a \textit{RF100} classifier trained on WatchAuth data. Marked points show thresholds. Observe that the true EER may lie between $ \sim 30\% $ and $ \sim 45\% $.}
	\label{fig:eerdiagramdegenerate}
\end{figure}

For models trained on limited data, many test samples are predicted the same probability, especially by the \textit{RF100} model. This poses a problem when calculating EER, as the crossover point between FAR and FRR can lie in a large interval (see Figure \ref{fig:eerdiagramdegenerate}). For reporting transparency, in Table \ref{tbl:DL_results_summary_limited} we present the EER as an interval, with the lower value representing the minimum value of FAR where FAR is greater than FRR, and the upper value representing the FRR value at this same threshold.

\begin{figure*}[t] 
	\centering
	\begin{subfigure}[b]{0.45\textwidth}
		\centering
		\includegraphics[width=\textwidth]{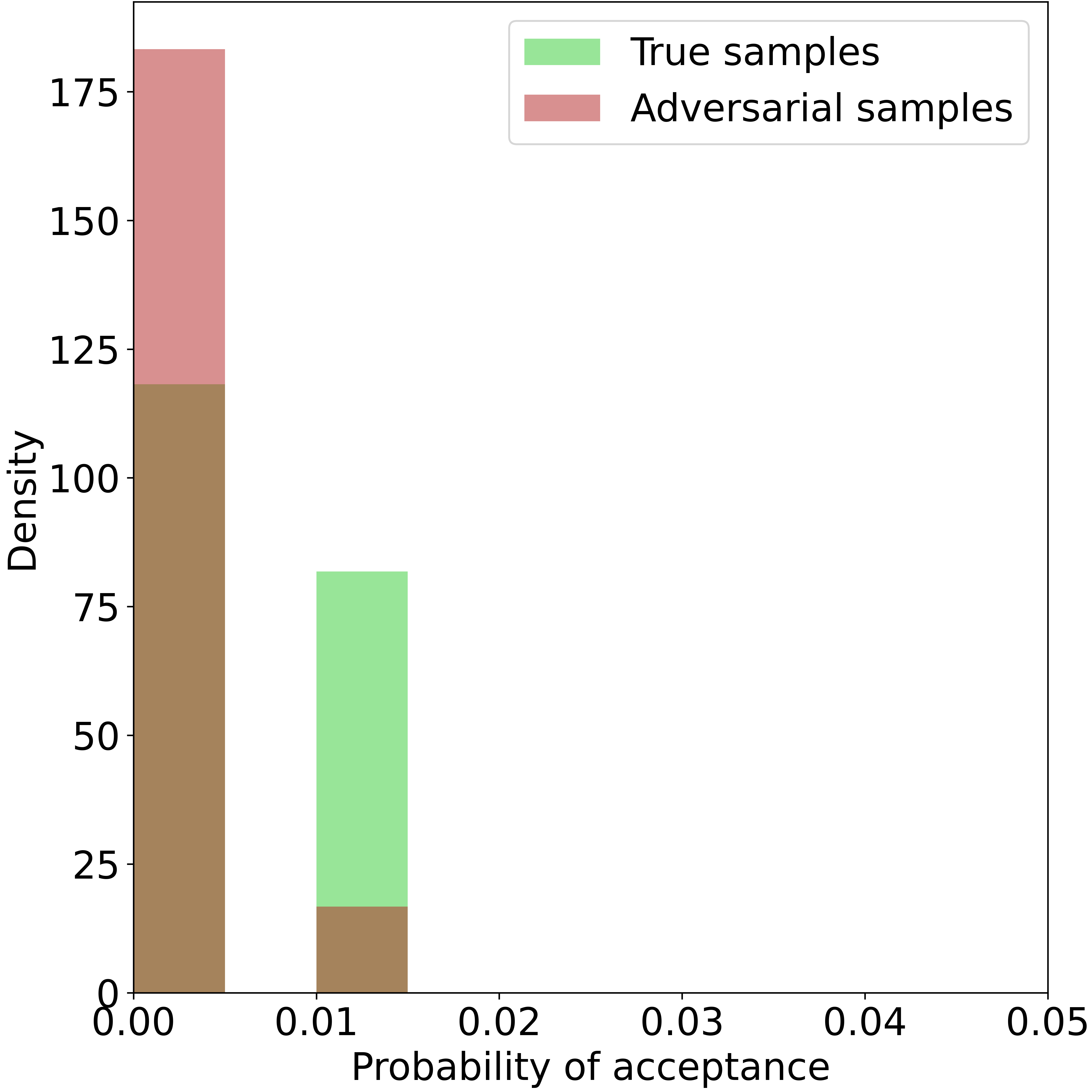}
		\caption[]%
		{{\small \textit{RF100}}}    
	\end{subfigure}
	\hfill
	\begin{subfigure}[b]{0.45\textwidth}
		\centering 
		\includegraphics[width=\textwidth]{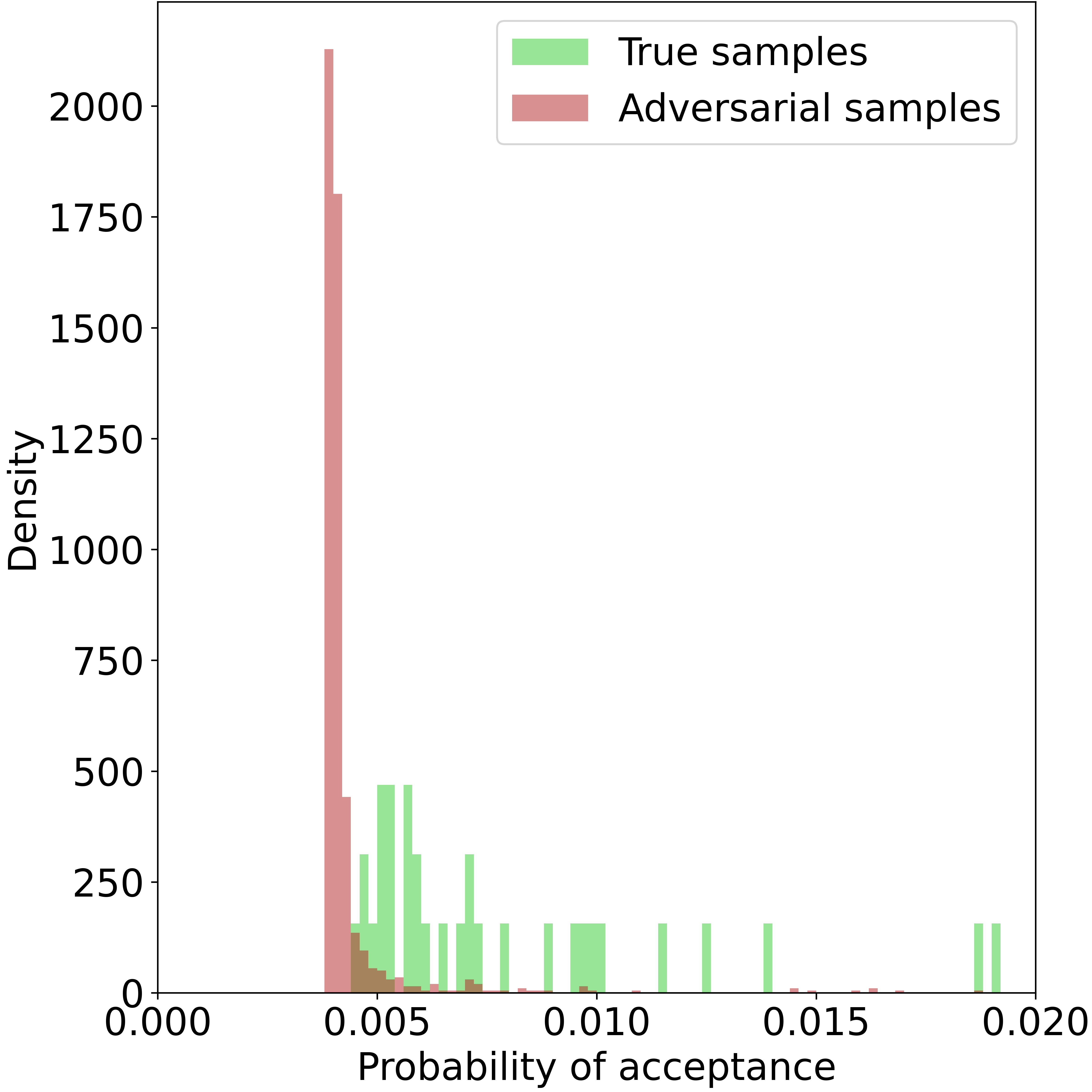}
		\caption[]%
		{{\small \textit{ComplexMix} }}    
	\end{subfigure}
	\hfill
	\caption[]%
	{{\small Distribution of acceptance probabilities by true and adversarial samples for classifiers trained to authenticate one user using limited training data.}}
	\label{fig:prob_dists_explained}
\end{figure*}

The random forest methods all performed very poorly on FAR@0 in our evaluation. Further analysis showed that the random forests were usually assigning some positive test samples zero probability (i.e. no decision tree classified them as a positive gesture), thus resulting in FAR@0 scores of $1.0$. In contrast, the deep-learned models all outperformed \textit{RF100} and \textit{RF1000} on FAR@0, suggesting they correctly assigned very low probability to a some negative samples without assigning very low probability to any positive samples. Figure \ref{fig:prob_dists_explained} visualises example probability distributions for \textit{RF100} and \textit{ComplexMix}, verifying this explanation.

When considering the setting of limited training data, the superiority of the \textit{ComplexMix} model is clear across all metrics.

\begin{table*}[h]
	\caption[]{\small\textit{Limited} experiment results for assorted models trained with only 10\% of a user's training and validation gestures.} 
	\label{tbl:DL_results_summary_limited}
	\vskip 0.15in
	\begin{center}
		\begin{sc}
			
\begin{tabularx}{\textwidth}{l|YYY} 
	\toprule
	Model       & AUROC  & EER interval   & FAR@0\\
	\midrule
	RF100       & 0.792  & (0.165, 0.339)  & 1.000  \\
	RF1000      & 0.858  & (0.192, 0.229)  & 0.904  \\
	\midrule
	MLP         & 0.818  & (0.240, 0.258) & 0.779   \\
	GRU         & 0.696  & (0.329, 0.359)  & 0.880 \\
    ConvNet     & 0.834  & (0.223, 0.243)  & 0.719  \\
    SimpleMix   & 0.803  & (0.229, 0.264)  & 0.815  \\
	ComplexMix  & 0.874  & (0.172, 0.198)  & 0.728 \\
	\bottomrule
\end{tabularx}

		\end{sc}
		\vskip -0.4in
	\end{center}
\end{table*}

\subsection{Comparison with published literature}
 
The EER value obtained for \textit{RF100} in Table \ref{tbl:DL_results_summary} is consistent with Sturgess et al., validating their key finding that payment gestures are a viable biometric for authentication. However, they do not note the issue with EER when data is scarce, so their results concerning limited data scenarios are misleading and not directly comparable to this work.

Furthermore, their FAR@0 statistics are at odds with those presented here. We noticed that their (publicly available) code used to calculated FAR@0 contained a bug which caused the published statistics to greatly over-represent the model's true ability in limited data scenarios. The bug arose when a decision threshold was set to 0 --- the FAR at that threshold was reported as the FAR@0 value, irrespective of the FRR at that threshold (which was often non-zero). Instead, we report a FAR@0 of 1.0 in the scenario where the FRR is never zero\footnote{For better stability, we actually check if the FRR is less than 0.01.}, which we believe better aligns with the behaviour we want FAR@0 to measure.

\section{Discussion}

While some models clearly outperformed others, all had similar performance to the \textit{MLP} baseline. This may suggest that despite the additional expressive power of the models considered, the majority of learnt features used to classify users were relatively simple.

Despite the authentication improvements they offer, the models trained in this chapter are unlikely to be suitable for deployment directly on a smartwatch due to memory and battery constraints. One possibility for creating more compact deep learning models is utilising the teacher-student framework \cite{hintonDistillingKnowledgeNeural2015}, where a pre-trained large model directly supervises the training of a smaller, more parameter-efficient network; this is left for future work.

The deep learning models in this chapter remain highly relevant for offline authentication-related tasks, for example retrospective fraud detection and user identification. Proving their effectiveness is also an essential building block for the work we conduct in Chapter \ref{ch:5-generative_modelling}.

\chapter{Generative modelling of payment gestures}\label{ch:5-generative_modelling}

\minitoc

\section{Introduction}

\subsection{Motivation}

At present, software run on a smartwatch must be memory-efficient and resource-sparse to work within limited memory constraints and battery capacity. Deep learning models are not generally sufficiently compact to adhere to these restrictions.

However, no such hardware limitations apply to the training phase for authentication models. This insight motivates a further direction of research considering whether large deep learning models can teach more compact models suitable for deployment on a smartwatch. We consider teaching in the form of generating additional data samples for training, as this approach is compatible with WatchAuth and could be used to reduce a user's enrolment burden.

The setting we find ourselves in is thus: given a large number of gestures from other users and a comparatively small number of gestures from a target user, how can we generate additional realistic gestures for the target user?

\subsection{Inadequacy of simple data transforms}

In contrast to other tasks such as image classification, it is not well understood which transformations preserve class label for authentication using gesture data. Common timeseries transformations such as jittering (adding Gaussian noise), temporal scaling and intensity scaling have been shown not to preserve class labels for user identification  \cite{beneguiAugmentNotAugment2020}. Furthermore, the axes of the watch sensors are fixed relative to the user's wrist, ruling out rotational data augmentations\footnote{This also rules out exploiting rotational symmetries in the model design (following the principles of Geometric Deep Learning).}.

\subsection{Deep learning approach}

Given the successes of Chapter \ref{ch:4-auth_improvements}, we may reasonably expect a deep learning model to have the capacity to achieve a stronger implicit understanding of which features are most discriminative between users than WatchAuth. To transfer this superior discriminative ability to a simpler system, we must be able to generate diverse synthetic gestures by conditioning on a specific user.


For this task, we considered the two most popular deep generative frameworks \cite{iglesiasDataAugmentationTechniques2023}, namely adversarial models such as GANs and autoencoder-based models such as VAEs. While GANs typically generate data of higher fidelity than autoencoder-based models, they frequently suffer from unstable training, particularly when data is scarce. This issue proved insurmountable in our preliminary investigations, and so attention was instead restricted to autoencoder-based models (see Section \ref{sec:3generativemodels}).

To use an autoencoder to generate synthetic data, we must first train the encoder and decoder together to embed points into a latent space and reconstruct gestures from their embeddings. After training, points are sampled from the learnt latent space and passed through the decoder to generate synthetic gestures.

To successfully generate user-specific gestures, an autoencoder model must satisfy the following two requirements:

\begin{enumerate}
	\item High reconstruction quality --- the decoder must reconstruct realistic gestures from embeddings in the latent space.
	\item Meaningful latent space representations --- the encoder must learn a latent space representation where nearby embeddings decode to semantically similar gestures (ideally gestures from the same user).
\end{enumerate}

\subsubsection{Architecture}\label{sec:5architecture}

We use the \textit{ComplexMix} model for our encoder architecture, given its success in 
Chapter \ref{ch:4-auth_improvements}.

\begin{figure}
	\centering
	\includegraphics[height=0.9\textheight]{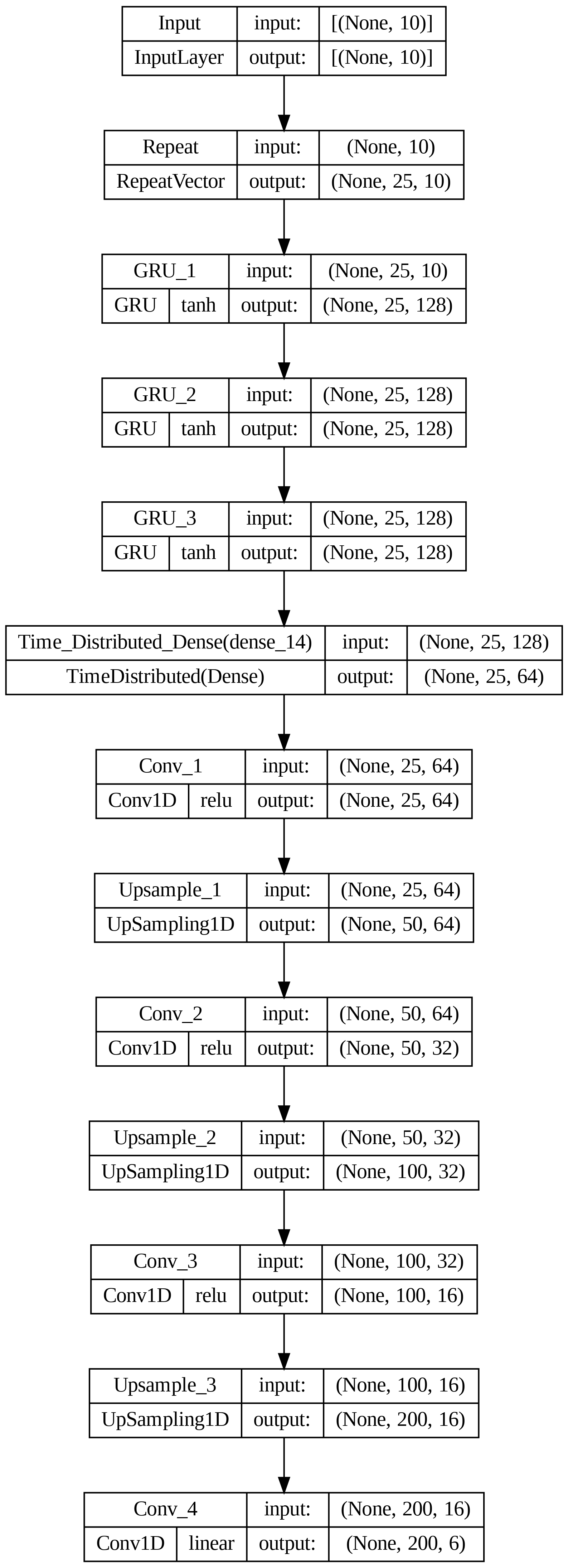}
	\caption{\textit{Decoder}}
	\label{fig:decoderarchitecture}
\end{figure}

Following a common theme in the literature (and validated empirically), we construct our decoder as an inverted version of the encoder. Our decoder consists of stacked GRU layers acting on the latent embedding vector, followed by a sequence of upsampling and convolutional layers to reconstruct a full-length output signal (see Figure \ref{fig:decoderarchitecture}).

The size of an autoencoder's latent space must be a compromise between the two requirements above, with larger spaces favouring reconstruction quality and smaller spaces favouring more meaningful latent space representations. Following extensive experimentation, the latent space size was fixed at 10-dimensional.

\section{Reconstructing high-quality gestures}

For synthetic data to be useful, it must be of sufficiently high quality to improve a classifier's performance on an authentication task. For an autoencoder, the quality of gestures is determined by the quality of the decoder's reconstructions. Given that the decoder is trained to minimise reconstruction loss, the choice of loss function strongly influences the quality of reconstructed gestures.

\subsection{Choice of loss function} \label{sec:5losses}

In the absence of a literature consensus on the choice of loss function for timeseries comparison, we investigate the effectiveness of several choices.

\subsubsection{Mean squared error}

A naive choice of loss is pointwise mean squared error (MSE). 

\[\qquad d_{MSE}(x,y) = \sum_{c=1}^{6} \sum_{n=1}^{200}  (x_{c,n} - y_{c,n})^2 \qquad .\]

MSE only captures the similarity between two signals in the intensity dimension.

\subsubsection{Soft-DTW}

Dynamic Time Warping (DTW) is an alternative quantity for measuring timeseries dissimilarity that is invariant to translation and warping in the time dimension (see Section \ref{sec:3DTW}). However, DTW is known to perform poorly as a loss function because the \textit{min} operation used to calculate it is not differentiable. In 2017, Cuturi \& Blondel introduced Soft-DTW \cite{cuturiSoftDTWDifferentiableLoss2017} as a smoothed version of DTW that replaces \textit{min} with a differentiable soft-min operation:

\[ \qquad  \text{soft-min}^{\gamma } (a_1, ..., a_n) = -\gamma \log \sum e^{-\frac{a_i}{\gamma}} \qquad .\]

The level of smoothing of the \textit{min} operation is controlled by $ \gamma $, which we set to a standard value of 0.1. We calculate Soft-DTW separately for each channel - otherwise, the time warping heavily penalises reconstructions that slightly misalign peaks for two separate channels. We use Maghoumi et al.'s open-source Cuda implementation for PyTorch \cite{maghoumiDeepNAGDeepNonAdversarial2021}, which follows the formulation of Soft-DTW given by Zhu et al. \cite{zhuDevelopingPatternDiscovery2018}.


\subsubsection{Keogh's lower bound for DTW}

Keogh's lower bound for DTW (see Section \ref{sec:3LB_Keogh}) is much more efficient to compute than Soft-DTW. Furthermore, it is tolerant of small translational shifts / warping within its Sakoe-Chiba bandwidth. However, it provides zero reconstruction loss when one curve is entirely within the envelope of another curve, which would limit an autoencoder's reconstruction resolution to the coarseness of the bandwidth.

We overcome this issue through a modification of Keogh's lower bound, which we name \textit{KLB-mod} and construct as the weighted average of Keogh's lower bound at exponentially varying bandwidths:

\[ \qquad d_{\text{\textit{KLB-mod}}}(x,y) = \sum_{w=1}^{5} (6-w) \cdot d_{KLB}^{(2^w)}(x,y) \qquad .\]
where $ d_{KLB}^{(w)}$ is as defined in Section \ref{sec:3LB_Keogh}. This formulation results in a theoretically smoother loss landscape than MSE or Keogh's lower bound with fixed bandwidth.

\subsubsection{Feature-based approaches}

WatchAuth compares gestures based on their feature similarity rather than on explicit differences between the original curve amplitudes. To reconstruct gestures that are highly discriminative for authentication, we consider augmenting the above reconstruction losses with a feature-based loss.

We represent a curve as a vector of differentiable features and apply a standard Euclidean kernel. The features used in WatchAuth that are compatible with TensorFlow are used for this feature vector representation, namely \textit{max}, \textit{min}, \textit{mean}, \textit{standard deviation}, \textit{variance}, \textit{skew}, \textit{kurtosis}, \textit{median} and \textit{inter-quartile range}.

We conducted experiments with loss functions defined as the weighted sum of this feature loss and each of the MSE and \textit{KLB-mod} loss functions, which we call \textit{MSE + Feature} and \textit{KLB-mod + Feature} respectively. The relative weight of the non-feature and feature losses was heuristically set to 0:0.1 and 1:0.01 respectively to reconstruct visually realistic gestures\footnote{We also attempted to calculate features on a rolling basis for a timeseries and compare two "feature timeseries". This approach resulted in noisy reconstructions and took prohibitively long to run.}.

\subsection{Experiments} \label{sec:5loss_experiments}

All models were implemented in TensorFlow \cite{martinabadiTensorFlowLargeScaleMachine2015}, with the exception of the model using the Soft-DTW loss function, which was implemented in PyTorch \cite{paszkePyTorchImperativeStyle2019} for compatibility with the Cuda implementation of Soft-DTW.

The same naive autoencoder (with architecture detailed in Section \ref{sec:5architecture}) was trained on the WatchAuth gesture data using different loss functions. Each model was trained until convergence with early stopping on its validation error, using the same random 20\% validation split for each loss.

The quality of reconstructions was evaluated quantitatively by the "train-synthetic test-real" method (TSTR). This is a strategy for evaluating the similarity between synthetic data and true data by training a classifier on the synthetic data and testing it on the real test dataset. Strong classification performance demonstrates explicitly that the synthetic data is of sufficiently high quality to be useful as training data. We used the TSTR method both with the \textit{ComplexMix} classifier and a reduced version of \textit{RF100} (limited to using just features which are compatible with TensorFlow) from Chapter \ref{ch:4-auth_improvements}.

We begin by using TSTR for classification of timeseries data into gesture and non-gesture classes, to investigate whether synthetic gestures were reliably distinguishable from non-gesture data. 240 synthetic gestures were generated as the direct reconstructions of randomly selected gestures for the positive class, with 240 non-gesture samples used as the negative class. This task is a relatively low bar to pass; the more interesting and difficult task of TSTR for authentication is carried out at a later stage in this report\footnote{TSTR for authentication requires training one model per user; this is very time-consuming so could not be done iteratively throughout the design process.}.

\subsubsection{On methods of evaluating synthetic data}

Alternative methods for evaluating the quality of synthetic data exist but are not as relevant as TSTR in this scenario. Fréchet Inception Distance (FID) is the image processing literature's choice of metric for evaluating the quality of synthetically generated images, and can be adapted to timeseries data. However, it only allows for \textit{relative} comparisons between data distributions and does not give insight into whether a synthetic dataset's distribution is adequately close to the real data to be useful as training data.


\subsection{Results}

Table \ref{tbl:reconstruction_results} summarises the AUROC and EER values computed for our TSTR intent recognition task.

\begin{table*}[h]
	\caption[]{\small TSTR intent results for autoencoders trained with different choices of reconstruction loss, using both \textit{ComplexMix} and \textit{RF100} as classifiers.} 
	\label{tbl:reconstruction_results}
	\vskip 0.15in
	\begin{center}
		\begin{sc}
			
\begin{tabularx}{\textwidth}{l|YYYY}
	\toprule
	& \multicolumn{2}{c}{ComplexMix} & \multicolumn{2}{c}{RF100}\\
	Loss function       & AUROC & EER & AUROC & EER \\
	\midrule
	No reconstruction   & 0.94 & 0.14 & 0.97 & 0.054  \\
	\midrule
	MSE                 & 0.88 & 0.19 & 0.97 & 0.054  \\
	Soft-DTW            & 0.30 & 0.65 & 0.99 & 0.033  \\
	KLB-mod             & 0.88 & 0.18 & 0.97 &  0.056 \\
	MSE + Feature       & 0.90 & 0.18 & 0.97 & 0.055  \\
	KLB-mod + Feature & 0.92 & 0.16 & 0.97 &  0.056 \\
	\bottomrule
\end{tabularx}

		\end{sc}
		\vskip -0.4in
	\end{center}
\end{table*}

\begin{figure}
	\centering
	\includegraphics[width=0.99\linewidth]{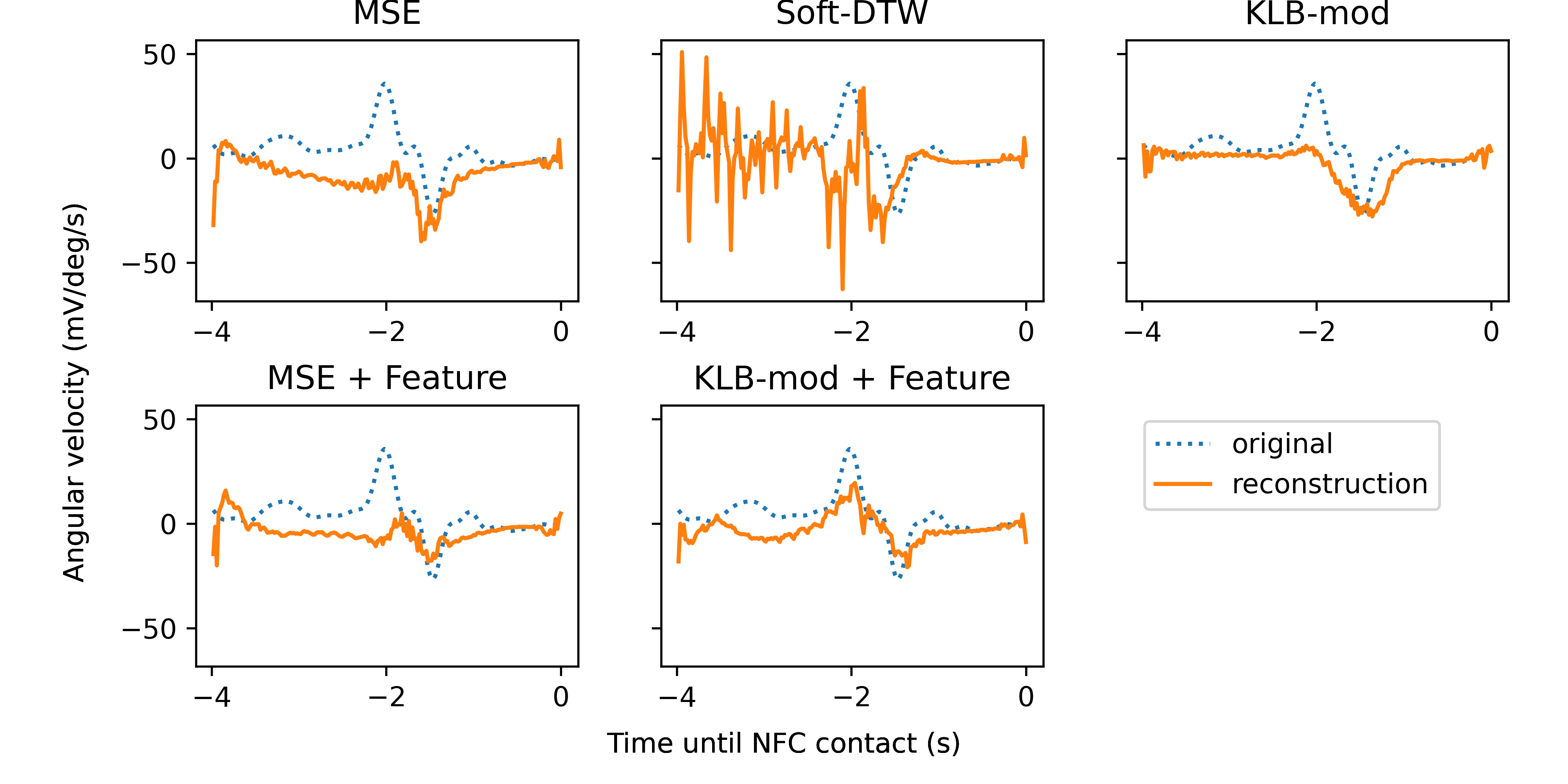}
	\caption{\small Visualising reconstructed gestures along the gyroscope z-axis for autoencoders trained with different loss functions.}
	\label{fig:gyrzlbkeoghstats}
\end{figure}

\subsection{Discussion}

We firstly note that models using all loss functions except Soft-DTW achieved strong results on both TSTR tasks. The Soft-DTW model generated samples that achieved superior intent recognition ability to the original curves on the \textit{RF100} TSTR task but completely failed the \textit{ComplexMix} TSTR task.

Visualising reconstructed gestures gives an explanation. We see in Figure \ref{fig:gyrzlbkeoghstats} that the reconstructed curve from the model using Soft-DTW is very noisy. This could be a result of difficulty training effectively --- gradients from the loss must be backpropagated through a dynamic programming algorithm, resulting in a highly complex optimisation landscape.

These reconstructed gestures are clearly extremely different to non-gesture data; perhaps this extreme difference to non-gesture data has resulted in the strong performance on the \textit{RF100} TSTR task despite the unrealistic generated gestures.

Of the other loss functions, \textit{KLB-mod + Feature} achieved the best AUROC and EER on the \textit{ComplexMix} TSTR task and strong performance on the \textit{RF100} TSTR task. We use this loss function going forwards.

\section{Regularising the latent space}\label{sec:5regls}

\begin{figure}[t]
	\centering
	\includegraphics[width=0.8\linewidth]{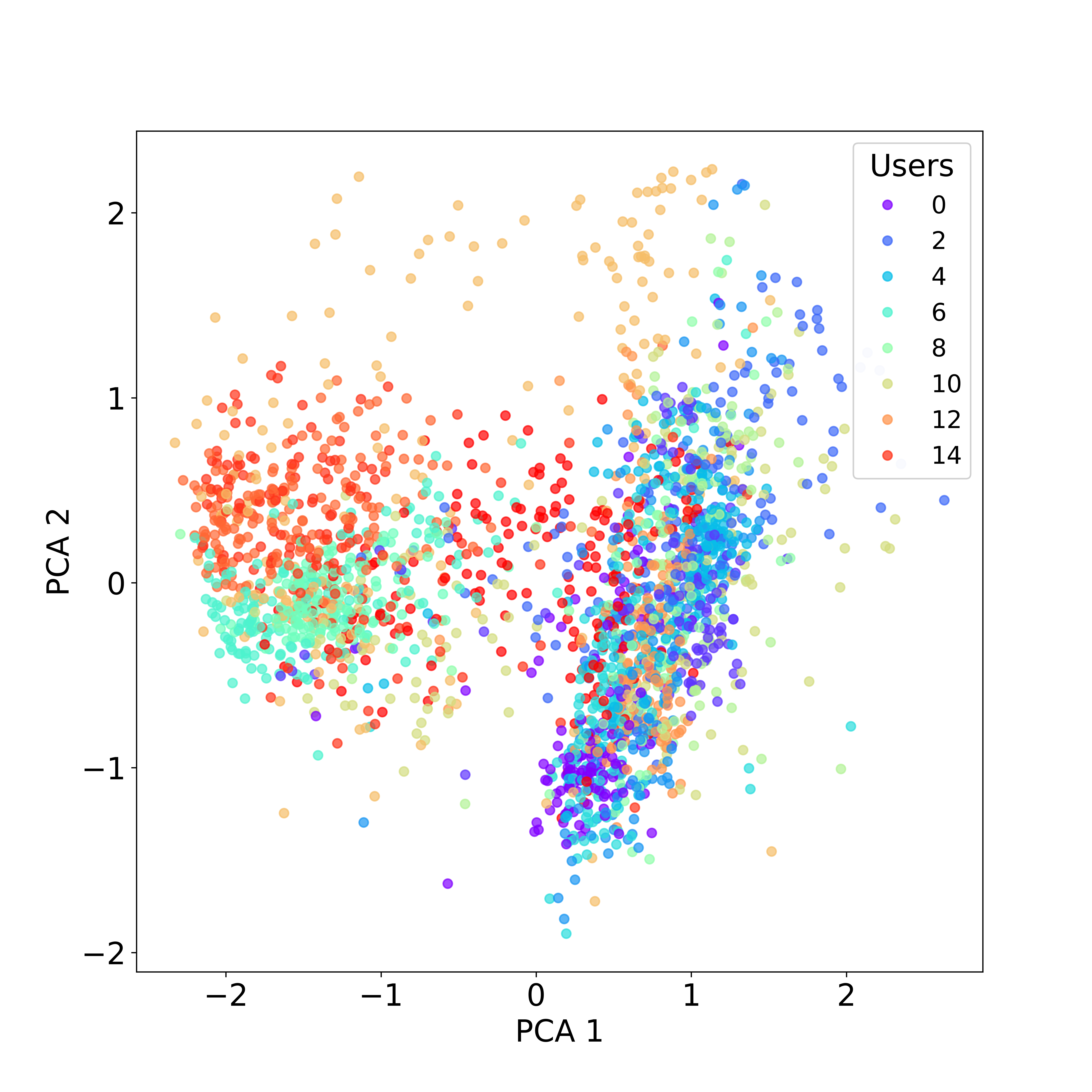}
	\caption{\small The first two principal components of the training data's latent space embedding for an unregularised autoencoder trained using the \textit{KLB-mod + Feature} loss.}
	\label{fig:latentspaceunregularised}
\end{figure}

Figure \ref{fig:latentspaceunregularised} visualises the latent space of an autoencoder model trained using \textit{KLB-mod + Feature}. 

It is promising that there appears to be a degree of natural clustering by user. However, the latent space lacks the necessary continuity to be useful for generating synthetic gestures (see Section \ref{sec:3generativemodels}) and so requires regularisation.

The most popular version of a regularised autoencoder in the literature is a Variational Autoencoder (VAE) (see Section \ref{sec:3vaes}).

\subsection{VAEs} \label{sec:5vaes}

The standard loss function for a VAE introduces a regularisation term $ \ell_{KL}$:

\[ \qquad \ell_{VAE} = \ell_{reconstruction} + \ell_{KL} \qquad .\]

\begin{figure}[t]
	\centering
	\includegraphics[width=0.8\linewidth]{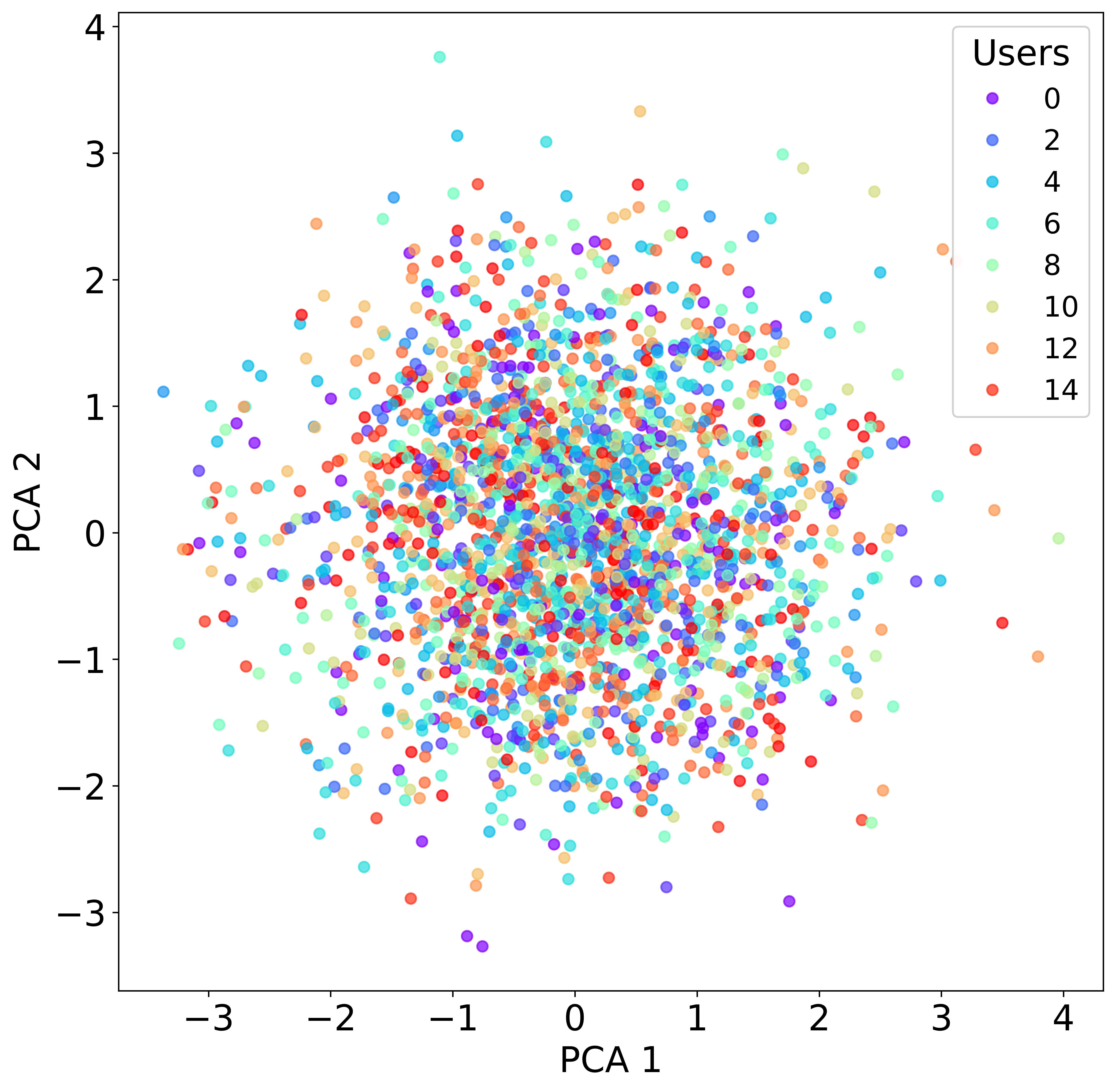}
	\caption{\small PCA visualisation of points sampled from the latent distributions of training data for an over-regularised VAE.}
	\label{fig:gaussianlatentspace}
\end{figure}

\begin{figure}[p]
	\centering
	\includegraphics[width=0.99\linewidth]{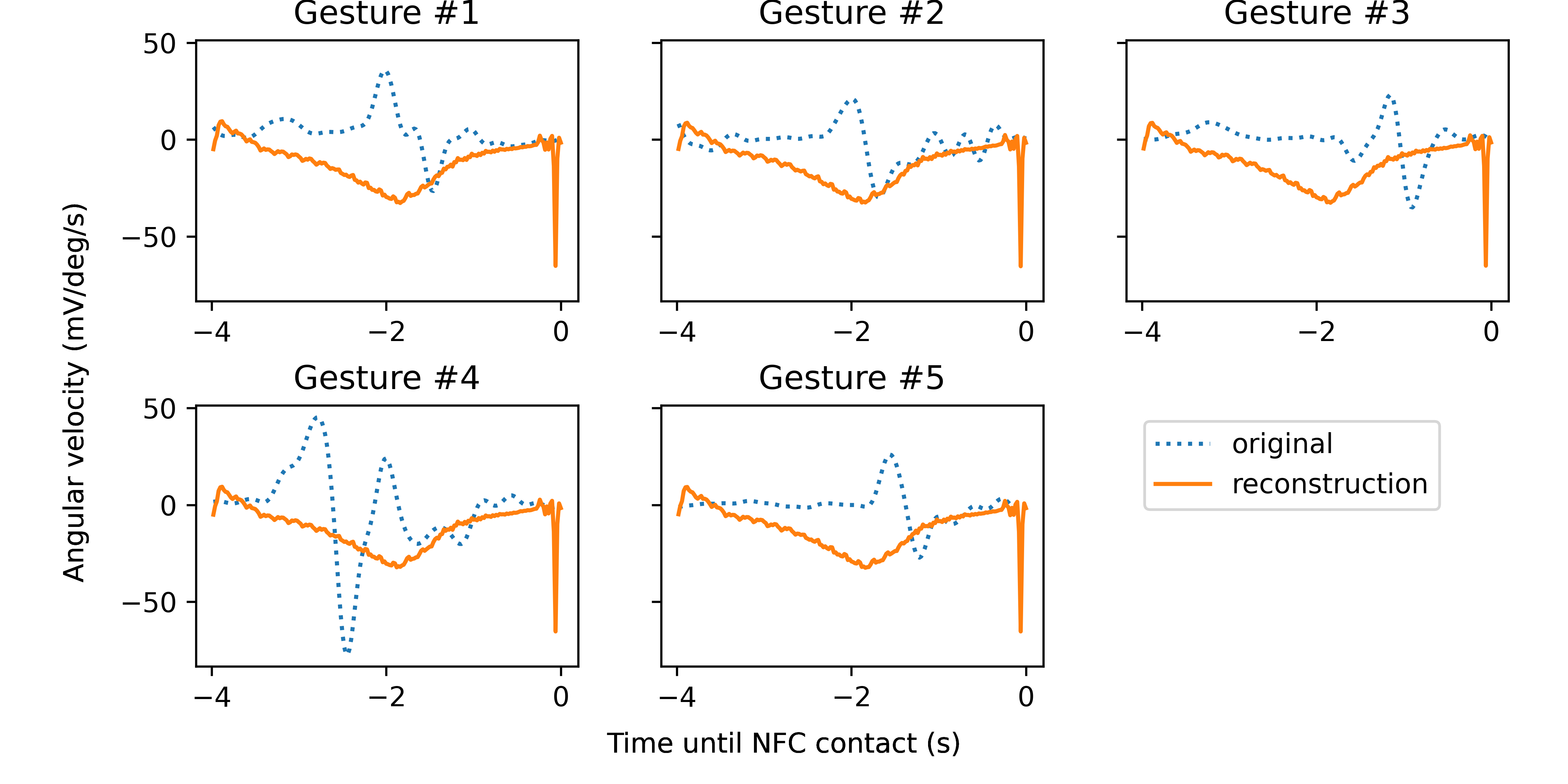}
	\caption{\small Reconstructed gestures (gyroscope z-axis shown) for an over-regularised VAE.}
	\label{fig:vaeoverregsamegestures}
\end{figure}

However, when we use this loss to train a VAE with the same underlying architecture as in Section \ref{sec:5architecture}, it is immediately apparent that the model is over-regularised. Figure \ref{fig:gaussianlatentspace} shows that the learned latent space appears to have standard Gaussian distribution, while Figure \ref{fig:vaeoverregsamegestures} shows that all gestures are reconstructed to the same curve.

To address this, we modify the balance between the KL loss and the reconstruction loss via a hyperparameter $ \beta $:

\[ \qquad  \ell_{VAE} = \ell_{reconstruction} + \beta \cdot \ell_{KL} \qquad .\]

\subsubsection{Selecting $ \beta $}

We exponentially varied the value of $\beta$ between 1 and $ 1e-6 $ and trained a VAE on the training dataset for each $ \beta $.

\begin{figure*}[p]
	\centering
	\begin{subfigure}[b]{0.99\textwidth}
		\centering
		\includegraphics[width=\textwidth]{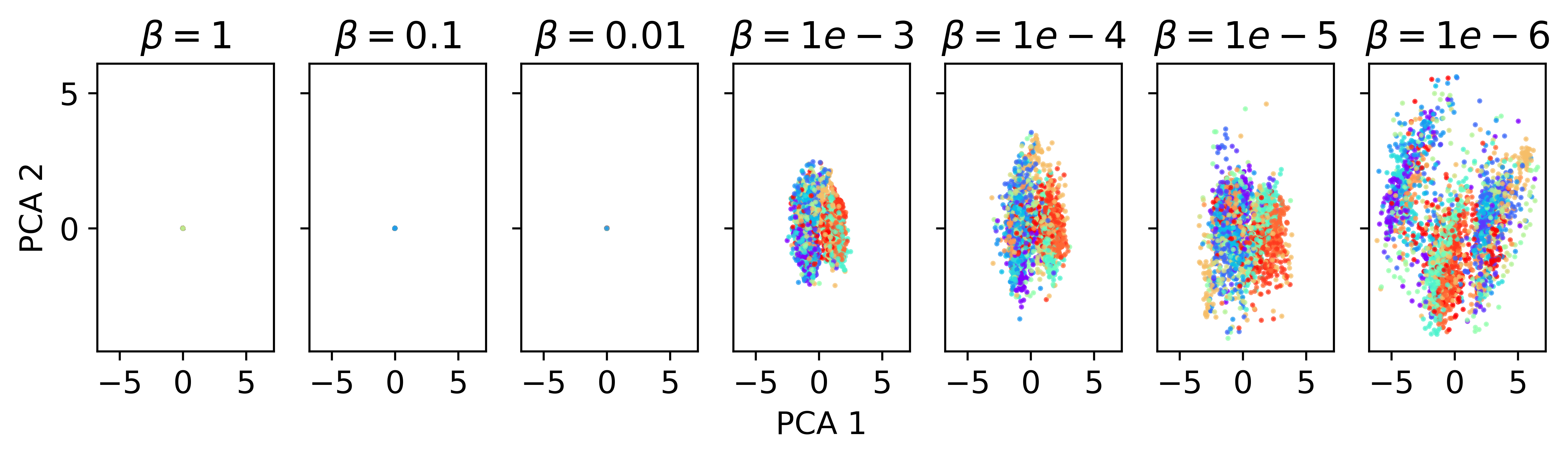}
		\caption[]%
		{{\small Embedding means}}
	\end{subfigure}
	\hfill
	\begin{subfigure}[b]{0.99\textwidth}
		\centering 
		\includegraphics[width=\textwidth]{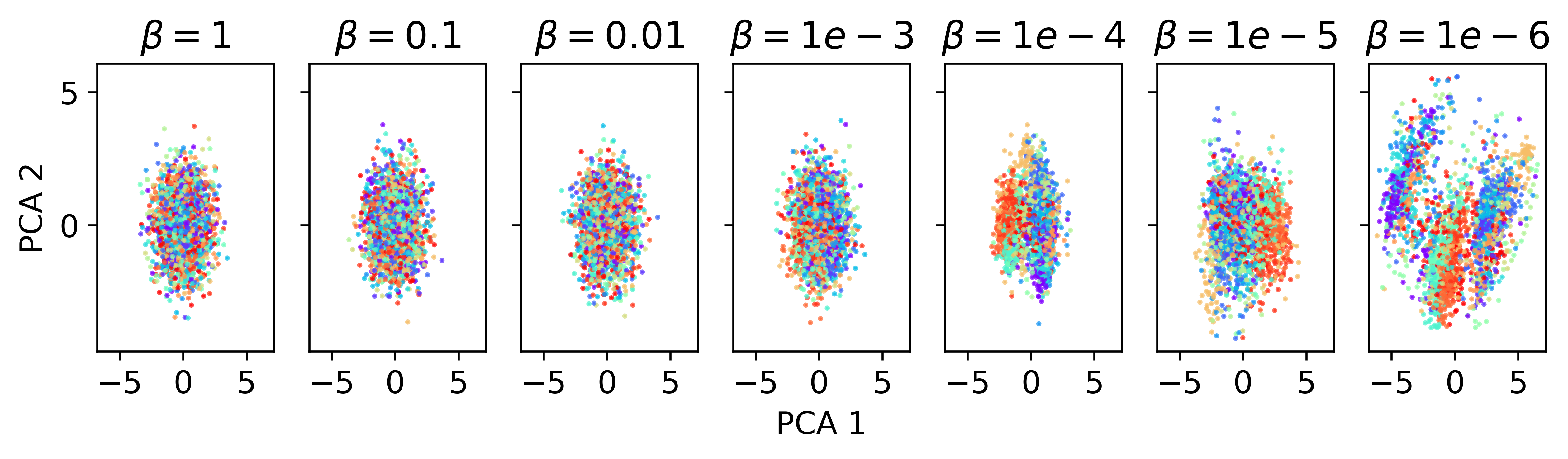}
		\caption[]%
		{{\small Sampled points}}
	\end{subfigure}
	\hfill
	\caption[]%
	{{\small Training data latent space embeddings for VAEs trained with decreasing $ \beta $ values. Different colours represent different users.}}
	\label{fig:beta_latent_spaces}
\end{figure*}

\begin{figure}
	\centering
	\includegraphics[width=0.9\linewidth]{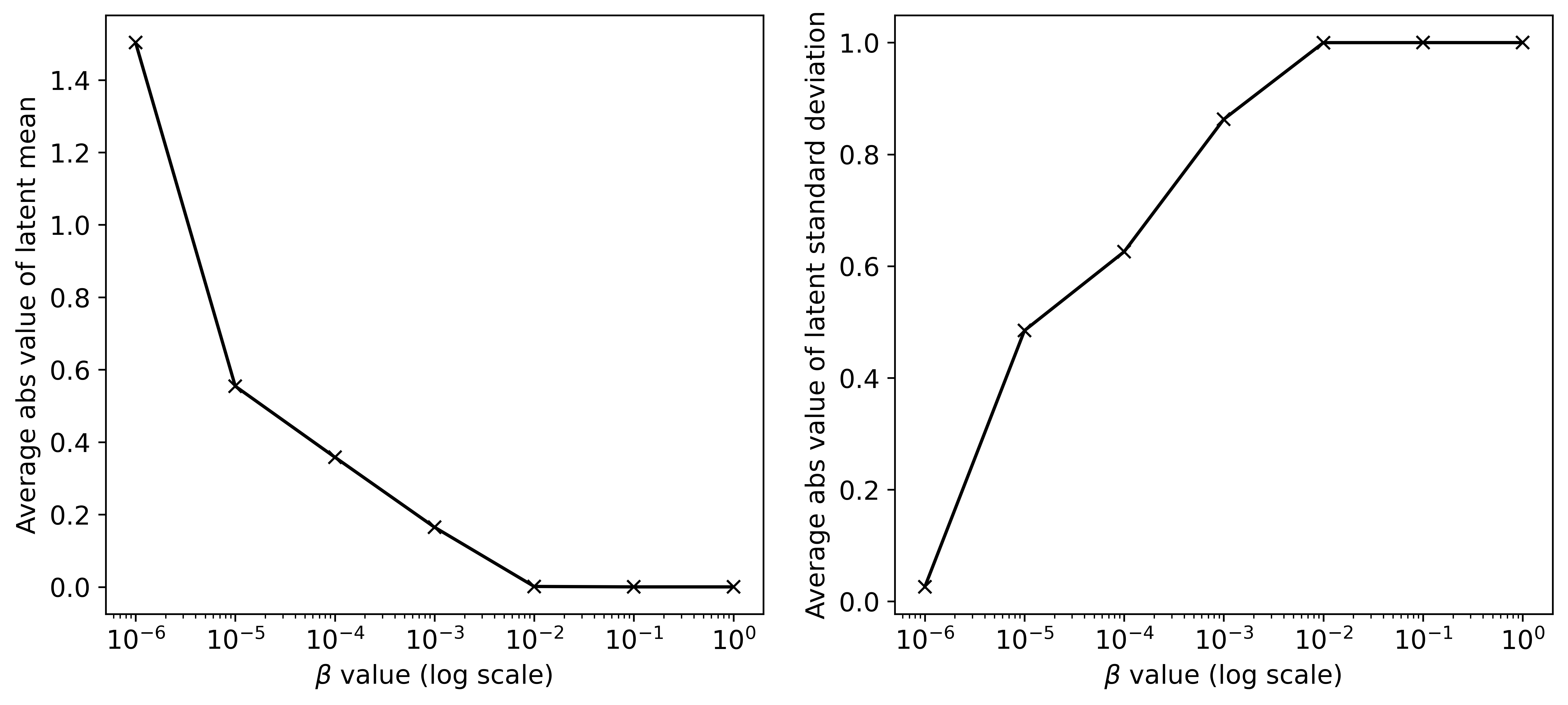}
	\caption{\small Average absolute values of latent embedding means (left) and standard deviations (right) for VAEs trained with varying $ \beta $.}
	\label{fig:valembeddingvaluesvaes}
\end{figure}

Figure \ref{fig:beta_latent_spaces} visualises how the latent space becomes less regularised as $ \beta $ decreases. We may quantify this by computing the average absolute values of embedding means and standard deviations, plotted in Figure \ref{fig:valembeddingvaluesvaes}. We can see that, for $ \beta \ge 1e-2 $, all points are encoded with approximate mean 0 and standard deviation 1. For $ \beta \le 1e-3$, regularisation still has an effect but is balanced with reconstruction ability.

\begin{figure}
	\centering
	\includegraphics[width=0.7\linewidth]{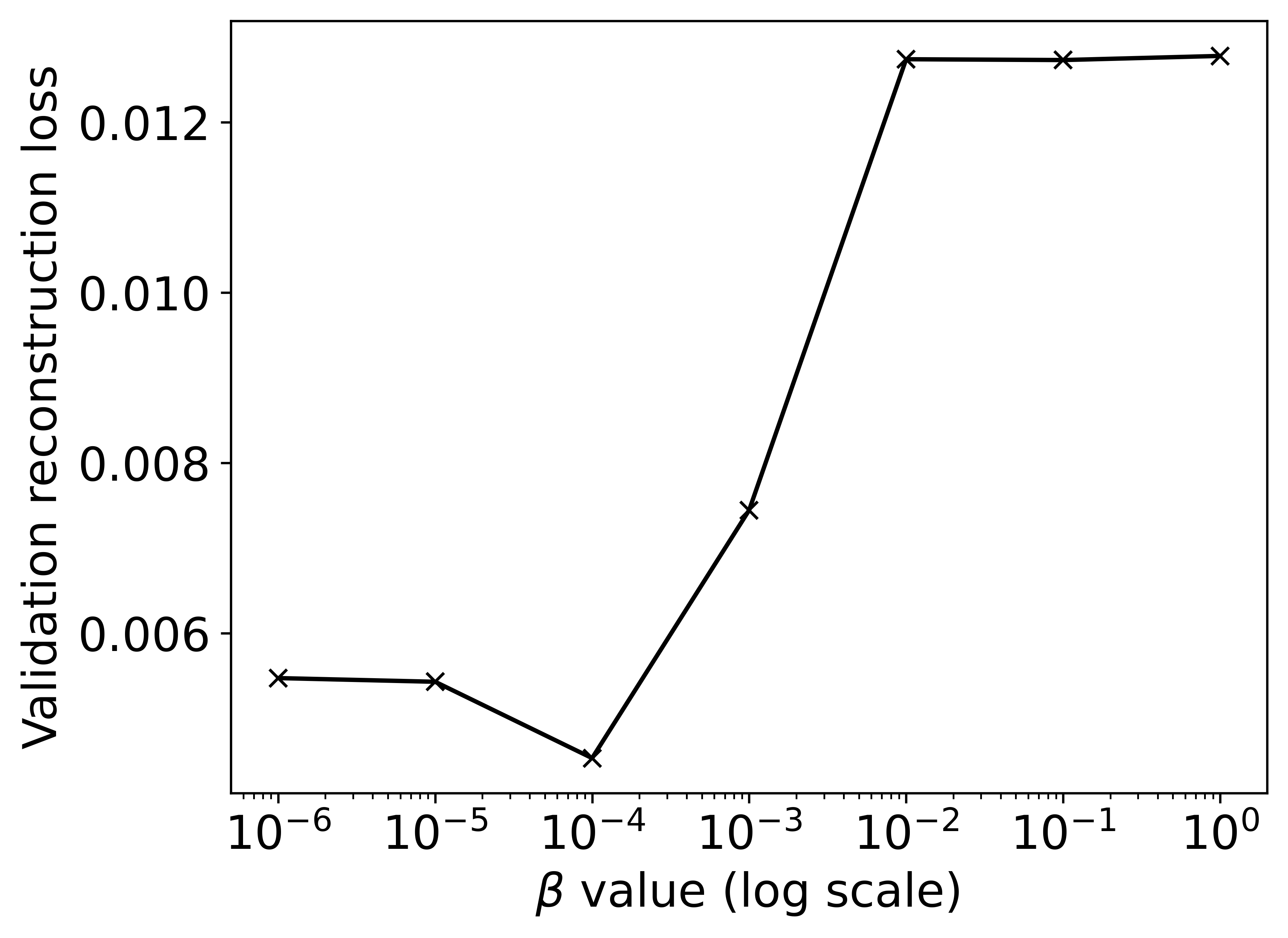}
	\caption{\small Validation reconstruction losses for VAEs trained with varying $ \beta $.}
	\label{fig:waesvallosses}
\end{figure}

Figure \ref{fig:waesvallosses} shows the validation reconstruction loss as $ \beta $ varies. Guided by this analysis, we chose $ \beta = 1e-4 $ to retain strong reconstruction ability while still regularising the latent space.

\subsection{WAEs} \label{sec:5waes}

WAEs are an alternative formulation of a regularised autoencoder that do not rely on a nondeterministic encoder for their regularisation loss term. In the literature, this has allowed them to achieve higher fidelity synthetic image reconstructions than VAEs. We consider whether this alternative regularisation scheme might result in higher quality reconstructions and a more meaningful latent space.

Similarly to Section \ref{sec:5vaes}, we balance the effect of the WAE regularisation loss term (see Section \ref{sec:3waes}) with a hyperparameter $ \beta $:

\[ \qquad \ell_{WAE} = \ell_{reconstruction} + \beta \cdot \ell_{regularisation} \qquad .\]

\subsubsection{Selecting $ \beta $}

WAEs exhibit similar behaviour to VAEs when $ \beta $ is set too high. As before, we varied $ \beta $ and balanced reconstruction and regularisation by choosing $ \beta = 1e-3 $ .

\subsection{Alternative training regimes}

We investigated pre-training the autoencoder model with $ \beta = 0 $ using non-gesture data. The concept behind this approach was akin to transfer learning, where we pre-train the model to have strong reconstruction ability before attempting to regularise the latent space.
An alternative approach we tried was gradually increasing the value of $ \beta $  during training. Neither approach had a measurable effect on the reconstruction quality or latent space behaviour.

\subsection{Verifying synthetic gestures via TSTR for intent}

\begin{table*}[h]
	\caption[]{\small TSTR intent results using \textit{ComplexMix}.} 
	\label{tbl:TSTR_intent_vaes_waes}
	\vskip 0.15in
	\begin{center}
		\begin{sc}
			
\begin{tabularx}{\textwidth}{l|YY}
	\toprule
	Regularisation    & AUROC & EER \\
	\midrule
	Unregularised     & 0.92 & 0.16 \\
	VAE               & 0.90 & 0.17 \\
	WAE               & 0.87 & 0.20 \\
	\bottomrule
\end{tabularx}

		\end{sc}
		\vskip -0.4in
	\end{center}
\end{table*}

We repeated the TSTR experiments of Section \ref{sec:5losses} using \textit{ComplexMix} with the VAE and WAE models to investigate whether adding regularisation affected the ability to reconstruct realistic gestures. These results are summarised in Table \ref{tbl:TSTR_intent_vaes_waes}.

We can see that the regularisation has had a slight effect, but that realistic gestures are still being reconstructed. WAEs performed worse than VAEs, so we proceed with the VAE regularisation architecture.

\section{Enforcing user-based clustering in the latent space}

To be able to sample points from the latent space and have confidence in their user label, we must achieve much more consistent clustering by user in the latent space than has been observed in the models so far. This motivates the explicit enforcement of user-based clustering in the latent space. 

To this end, we add a simple MLP classifier to our VAE model, and train it to classify points in the latent space according to their user. We strictly limit the expressive power of this classifier. As a result, to improve the latent space user classification the encoder must learn to organise the latent space by user.

Under the hypothesis that the distribution of gestures is determined both by user and non-user variables (e.g. terminal position, independent random effects), we only apply this classifier to the first 5 dimensions of the 10-dimensional latent space.

The natural loss function for this classifier is categorical cross-entropy. Empirically, we observed that this loss caused the model to increase its confidence in points it was correctly classifying rather than focussing on classifying all points correctly. An alternative choice of loss that we found to induce stronger clustering was a rank-based loss function, where points in the latent space are assigned scores per user, and the objective is to maximise the mean reciprocal rank (MRR) of the true user (thereby rewarding higher predicted ranks for the true user).

For differentiability and compatibility with our deep learning architecture, we used the approximate MRR proposed by Qin et al. \cite{qinGeneralApproximationFramework2010} and implemented in TensorFlow Ranking \cite{pasumarthiTFRankingScalableTensorFlow2019}. For an array of true labels $ \{y\} $ and predictions $ \{s\} $ (with a constant value for temperature):

\[\qquad  \ell_{Approx MRR}(\{y\}, \{s\}) = -\sum_{i} \frac{1}{\text{approxrank}_i} y_i \qquad ,\]
where 
\[\qquad  \text{approxrank}_i = 1 + \sum_{j \neq i}
\frac{1}{1 + \exp\left(\frac{-(s_j - s_i)}{\text{temperature} }\right)} \qquad .\]

This latent space authentication loss gives rise to a new loss function for our VAE:

\[ \qquad \ell_{total} = \ell_{reconstruction} + \beta \cdot \ell_{KL} + \alpha \cdot \ell_{auth} \qquad .\]

\begin{figure}
	\centering
	\includegraphics[width=0.9\linewidth]{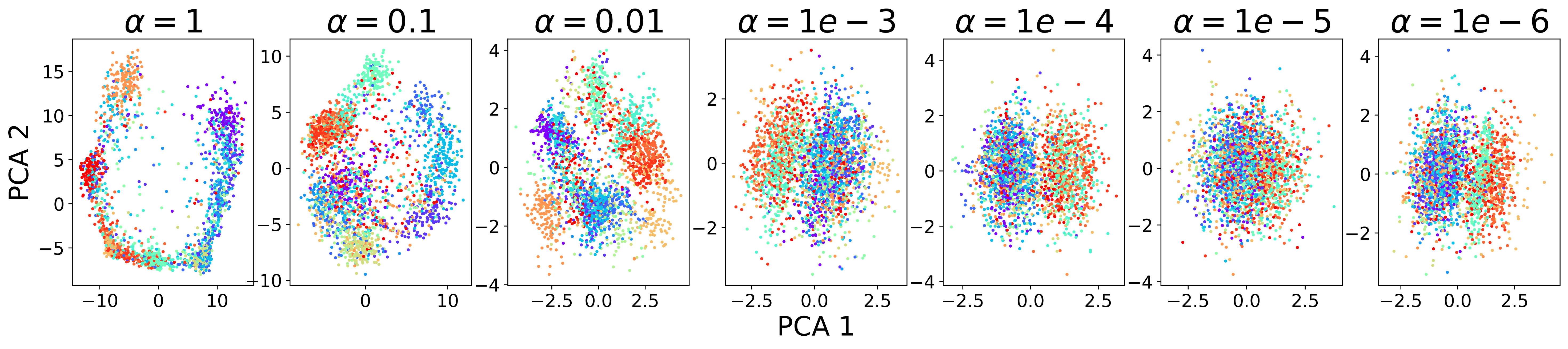}
	\caption{\small Latent space visualisation for VAEs trained with for varying values of $ \alpha$.}
	\label{fig:alphazs}
\end{figure}

We introduce a new hyperparameter $ \alpha $ to control the relative effect of the authentication loss. As in Section \ref{sec:5regls}, we experimented with different values of $ \alpha $ to balance reconstruction quality, regularisation and classification by user. Figure \ref{fig:alphazs} shows the latent space at varying values of $ \alpha $ --- setting $ \alpha $ too high results in excellent grouping by user and very poor regularity.

\begin{figure}[t]
	\centering
	\includegraphics[width=0.7\linewidth]{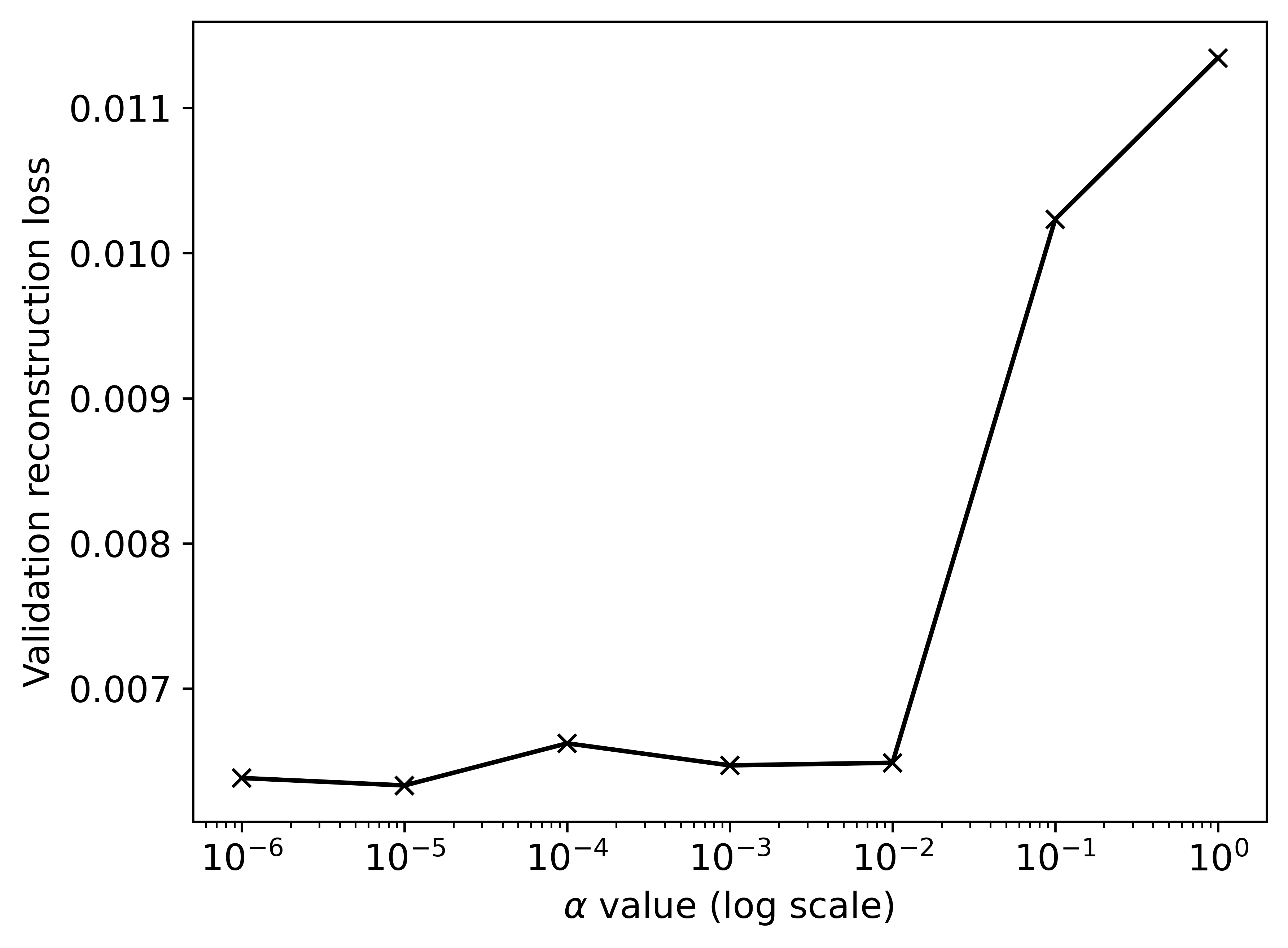}
	\caption{\small Validation reconstruction loss for VAEs trained with for varying values of $ \alpha$.}
	\label{fig:valreconlossauthvaes}
\end{figure}

Guided by the effect on reconstruction loss shown in Figure \ref{fig:valreconlossauthvaes}, we set $ \alpha = 1e-2 $. The validation approximate MRR is $ \approx 0.1$ without and $ \approx 0.3$ with the authentication MLP, which shows it induces an improvement in clustering by user.

\pagebreak
\section{Verifying model properties}

\subsection{Latent space structure}

\begin{figure}[b]
	\centering
	\includegraphics[width=0.7\linewidth]{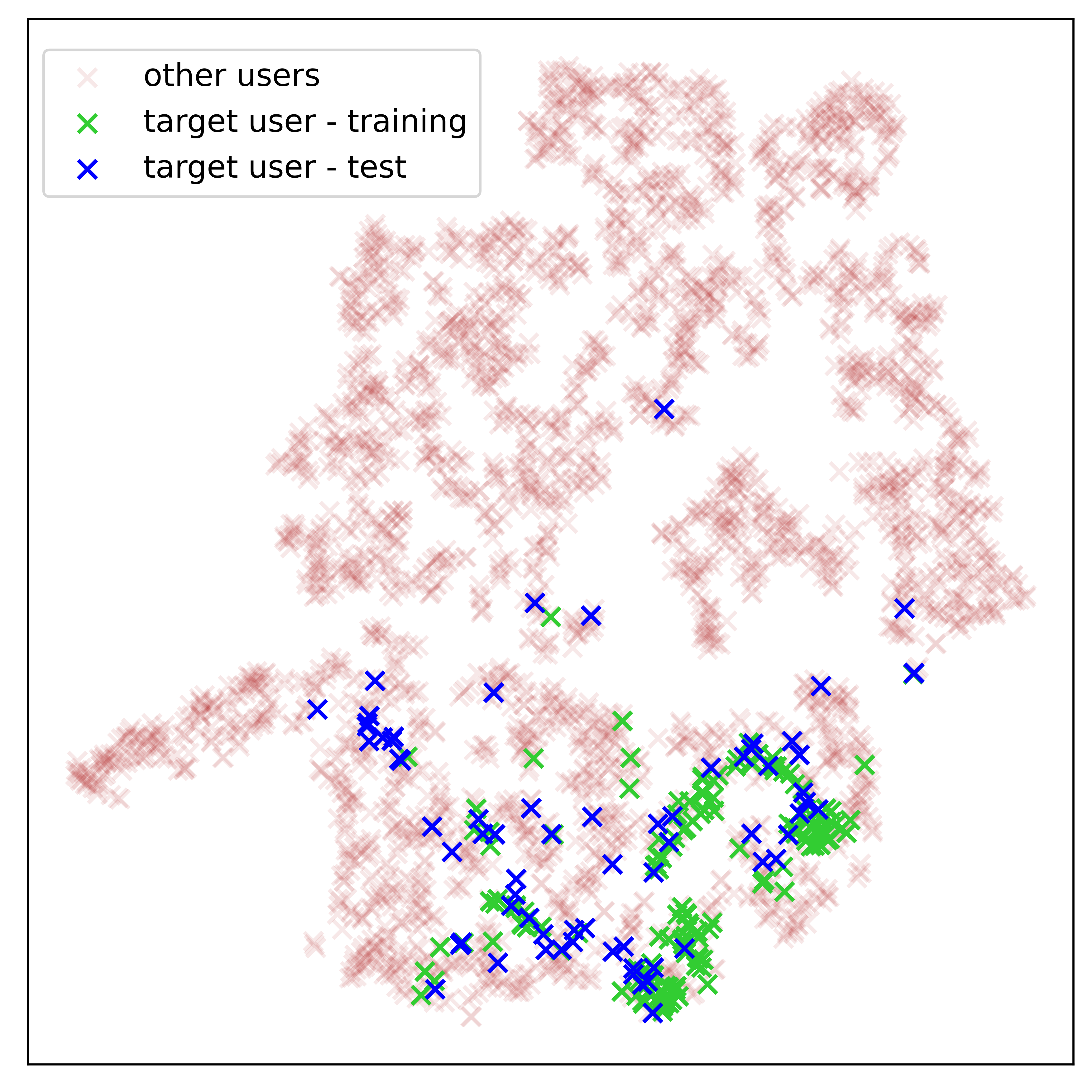}
	\caption{\small t-distributed Stochastic Neighbour Embedding (t-SNE) visualisation of a trained VAE's latent space, demonstrating that an example user's test gestures are embedded similarly to the user's training gestures.}
	\label{fig:unseenembeddings}
\end{figure}

For synthetic gestures to be useful, our encoder must assign similar embeddings to training gestures and unseen gestures for a given user. Figure \ref{fig:unseenembeddings} shows the result of embedding one user's gestures in the test dataset for a model trained using gestures in the training dataset.

It may be observed that the gestures observe a degree of localisation to a region of the embedding manifold. It is unrealistic to expect perfect clustering given that, in Chapter \ref{ch:4-auth_improvements}, our classifiers did not achieve perfect classification accuracy. With more users, the quality of gesture clustering would be expected to improve.

\subsection{Diversity of synthetic gestures}

\begin{figure}[t]
	\centering
	\includegraphics[width=0.7\linewidth]{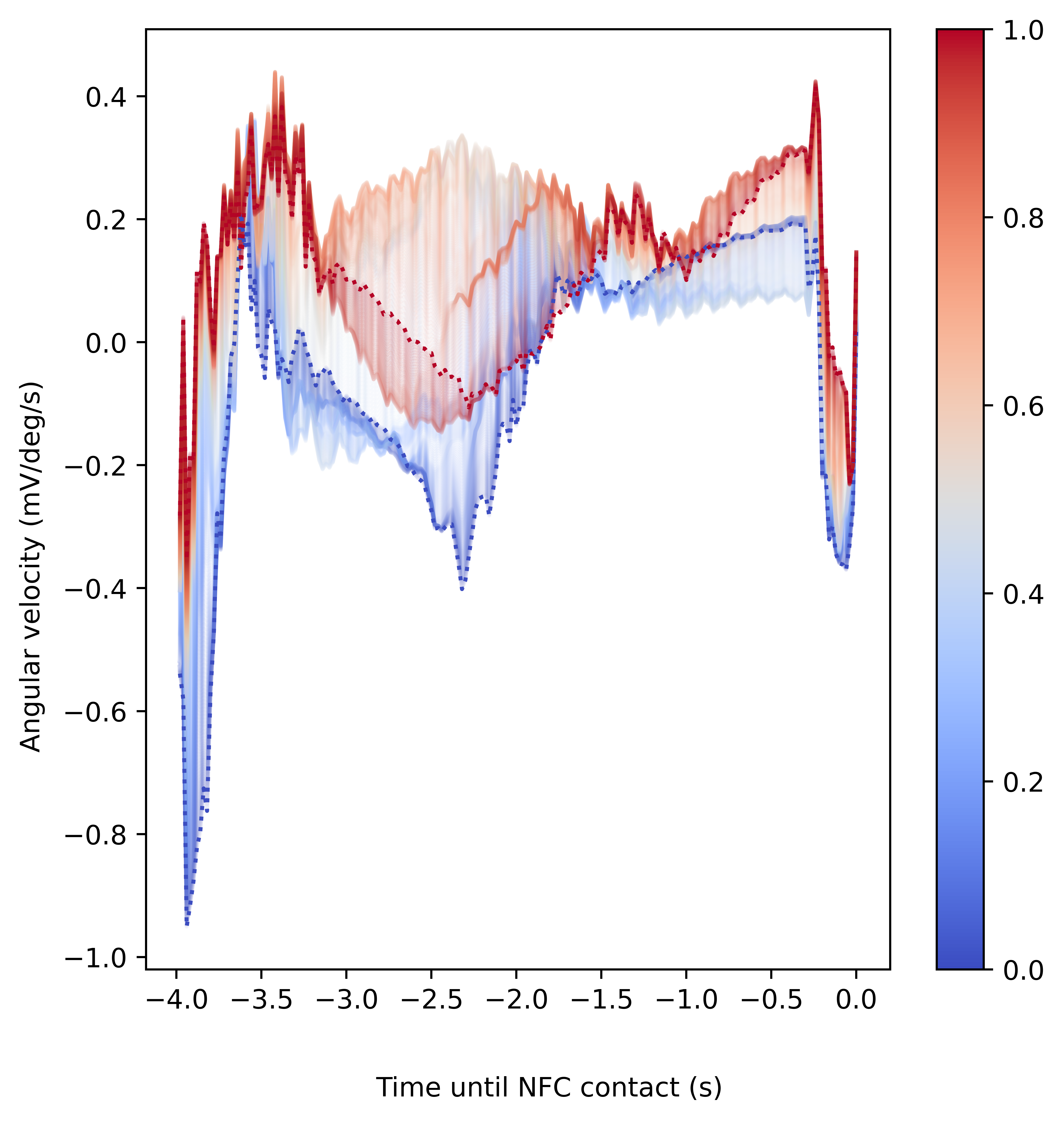}
	\caption{\small Visualisation of decoding points along a line segment in the latent space. For points $ x $ and $ y $, we visualise the decoding from latent space point $ \mu \cdot x + (1-\mu) \cdot y $ for each $ 0 \le \mu \le 1$. The reconstructions with $ \mu = 0 $ and $ \mu = 1 $ are represented by blue and red respectively. Colours associated with other values of $ \mu $ are shown by the colour bar.}
	\label{fig:notlinear}
\end{figure}

We verify that the decoder is generating diverse samples and is not just linearly interpolating between existing gestures. This can be seen in Figure \ref{fig:notlinear}, which we created by sampling two nearby points from the latent space, calculating points on the line between them, and decoding these intermediate points.

\chapter{Synthetic data for training WatchAuth}\label{ch:6-evaluating_gestures}

\minitoc

\section{Methodology}

\subsection{Generating synthetic gestures for authentication}

Suppose we have trained our VAE model from Chapter \ref{ch:5-generative_modelling} on a large volume of existing user data, and consider the setting where we have collected some gestures from a new user already. We wish to generate synthetic gestures for the new user using our trained model. As our trained model has not been exposed to the new user, we rely on the assumption that given a large enough corpus of users we may find a latent space that adequately captures common discriminative features between users.

We can apply the encoder to embed the target user's gestures into the VAE model's latent space and use these embeddings to guide the selection of further points in the latent space, which we then decode to generate our synthetic data. We investigate the effectiveness of four simple strategies for selecting points in the latent space.

\subsection{Latent space sampling strategies}

\subsubsection{Neighbourhood sampling}

We simply sample points using the embedding means and variances for the points we already know for the user.

\subsubsection{Self-mixed sampling}

We pick several of the existing embedding means for the user, and generate a new point in the latent space by calculating a randomly weighted convex combination of these points.

\subsubsection{Adversarial sampling}

We randomly pick two embeddings: one from our target user and one from a different user. We then calculate a convex combination of them weighted heavily towards the target user (85:15). The rationale behind this strategy is to generate gestures that should be classed as the target user but have some characteristics of other users, to generate a more diverse training set.

\subsubsection{Same-user sampling}

We take existing embeddings for the target user. We assume that as the authentication MLP is trained on the first 5 dimensions of the latent space that these dimensions are user-specific and other dimensions are not. We therefore set the embedding's last 5 dimensions to randomly sampled embeddings for other users' points.

\subsubsection{Other approaches}
More sophisticated strategies are an active topic of research. For example, Chadebec \& Allassonnière propose a geometry-aware random walk strategy for sampling the latent space \cite{chadebecDataAugmentationVariational2021}. This level of sophistication was not considered necessary in the current work, and is left for future extensions.

\subsection{Experiments}

To evaluate whether our methodology can improve the classification performance of a simple authentication model, we performed TSTR for authentication using the same version of \textit{RF100} as in Chapter \ref{ch:5-generative_modelling}. We trained 16 models, leaving out training data for one user each time. This mimics the realistic scenario of pre-training a generative model before seeking to enrol a new user into an authentication system with the aid of synthetic gestures.

For each trained model, we allowed access to two real user gestures per terminal position (14 in total), which were used to generate 500 synthetic gestures with each of our four sampling strategies. The original and synthetic gestures together formed the positive class for training an authentication classifier. For the results in Sections \ref{sec:6overall} and \ref{sec:6peruser}, our negative class contained the VAE's reconstruction of each of the other users' gestures.
We then converted our training data (using both real and synthetic gestures) into feature vectors and trained a random forest classifier for authentication (using analogous methodology to that in Section \ref{sec:5loss_experiments}).

In Section \ref{sec:6reduceburden}, we also included the original gestures from other users as negative samples, and varied the number of real user gesture samples per terminal to investigate how effective our methodology is for reducing enrolment burden.

\section{Results}

\subsection{Overall performance} \label{sec:6overall}

TSTR results for authentication are presented in Table \ref{tbl:TSTR_auth}.

\begin{table*}[h]
	\caption[]{\small TSTR authentication results for a random forest classifier aided by synthetic data generated by different strategies.} 
	\label{tbl:TSTR_auth}
	\vskip 0.15in
	\begin{center}
		\begin{sc}
			
\begin{tabularx}{\textwidth}{l|YYY}
	\toprule
	Sampling Strategy & AUROC & EER interval & FAR \\
	\midrule
	Original data only & 0.83 & (0.14, 0.28) & 1.00 \\
	\midrule
	Neighbourhood     & 0.83 & (0.23,0.26) & 0.79 \\
	Self-mixed        & 0.82 & (0.23,0.27) & 0.78 \\
	Adversarial       & 0.84 & (0.21,0.25) & 0.74 \\
	Same-user         & 0.82 & (0.23,0.27) & 0.77 \\
	\bottomrule
\end{tabularx}

		\end{sc}
		\vskip -0.4in
	\end{center}
\end{table*}

We cannot conclude that the addition of the synthetic data had a significant effect on the magnitude of AUROC and EER, although it did reduce the uncertainty around the true EER value.

The synthetic data provided a significant improvement to FAR@0, irrespective of sampling strategy. Adversarial sampling yielded the strongest performance on all metrics amongst the sampling strategies.

To investigate the cause of this improvement it is necessary to consider performance on a per-user basis.

\subsection{Per-user performance} \label{sec:6peruser}

\begin{figure}[p]
	\centering
	\includegraphics[width=0.9\linewidth]{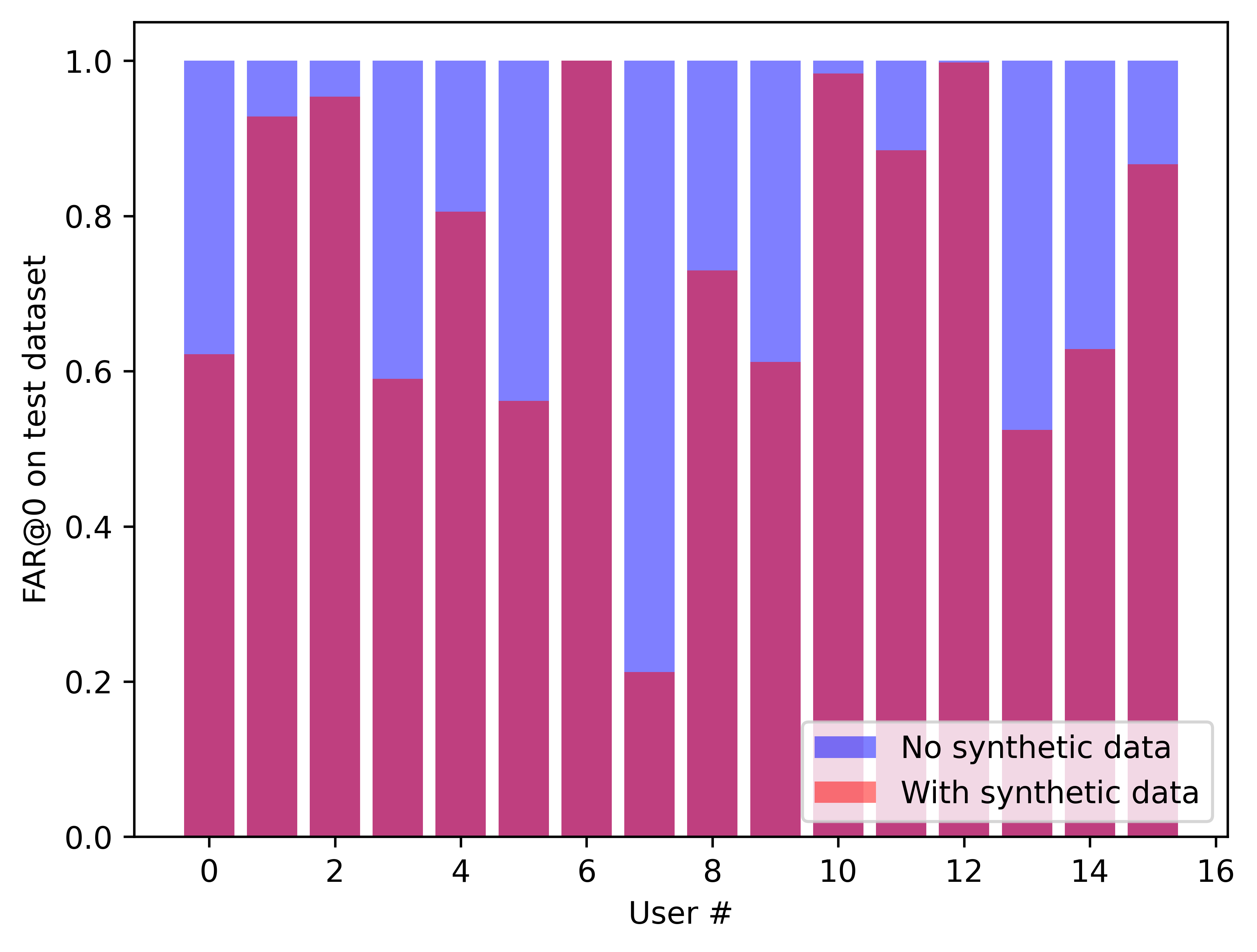}
	\caption{\small Average FAR@0 for random forest classifiers, trained with and without synthetic gestures (generated using adversarial sampling). Results are broken down by user.}
	\label{fig:farforadvsampling}
\end{figure}

Figure \ref{fig:farforadvsampling} shows the FAR@0 for each user when synthetic gestures were generated using adversarial sampling. It is clear that performance varied widely between users. For example, user 7 experienced a reduction in FAR@0 from $ 1.0 $ to $\sim 0.2 $, while users 6, 10 and 12 had very little or no reduction.

\begin{figure*}[p] 
	\centering
	\begin{subfigure}[b]{0.475\textwidth}
		\centering
		\includegraphics[width=\textwidth]{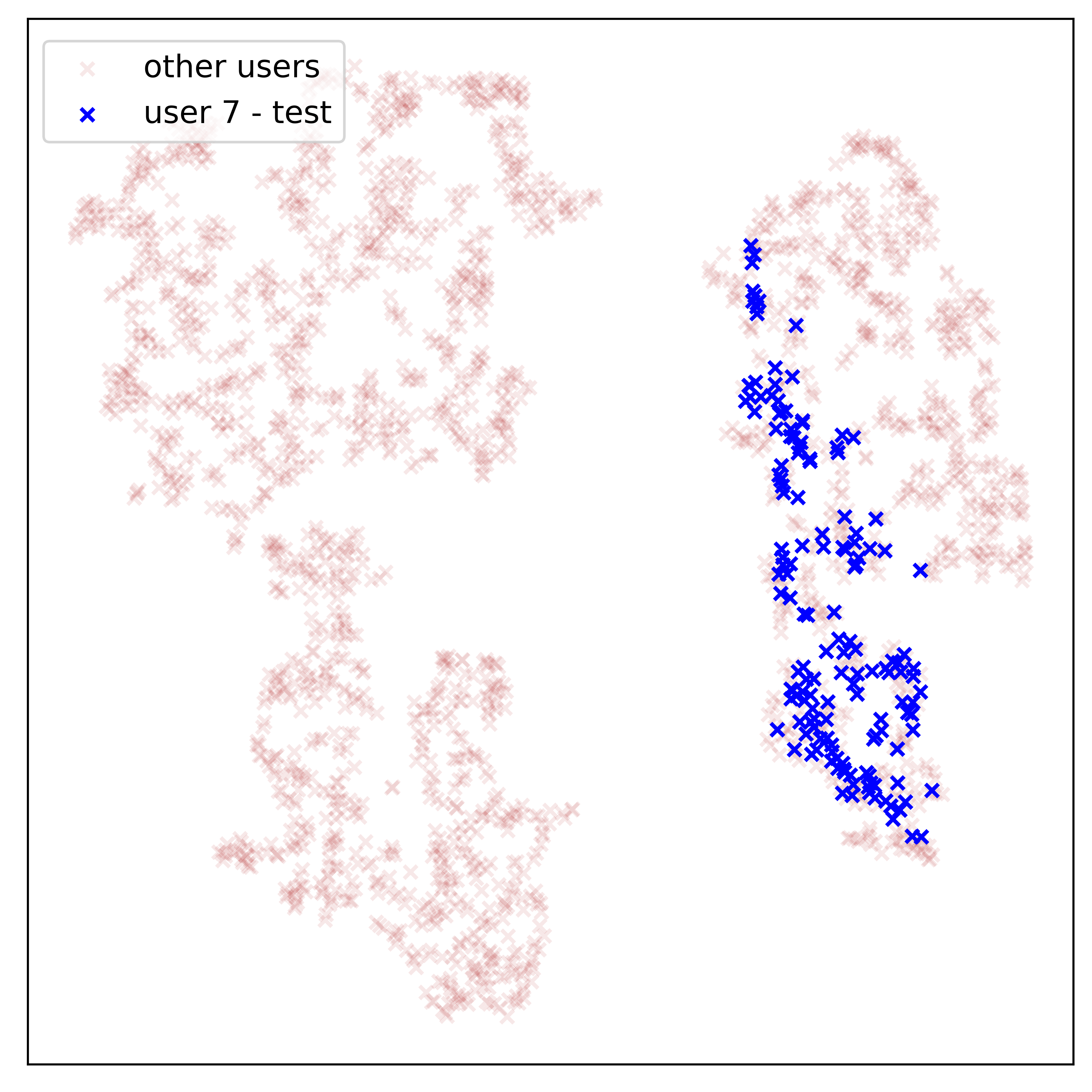}
		\caption[]%
		{{\small User 7}}
	\end{subfigure}
	\hfill
	\begin{subfigure}[b]{0.475\textwidth}
		\centering 
		\includegraphics[width=\textwidth]{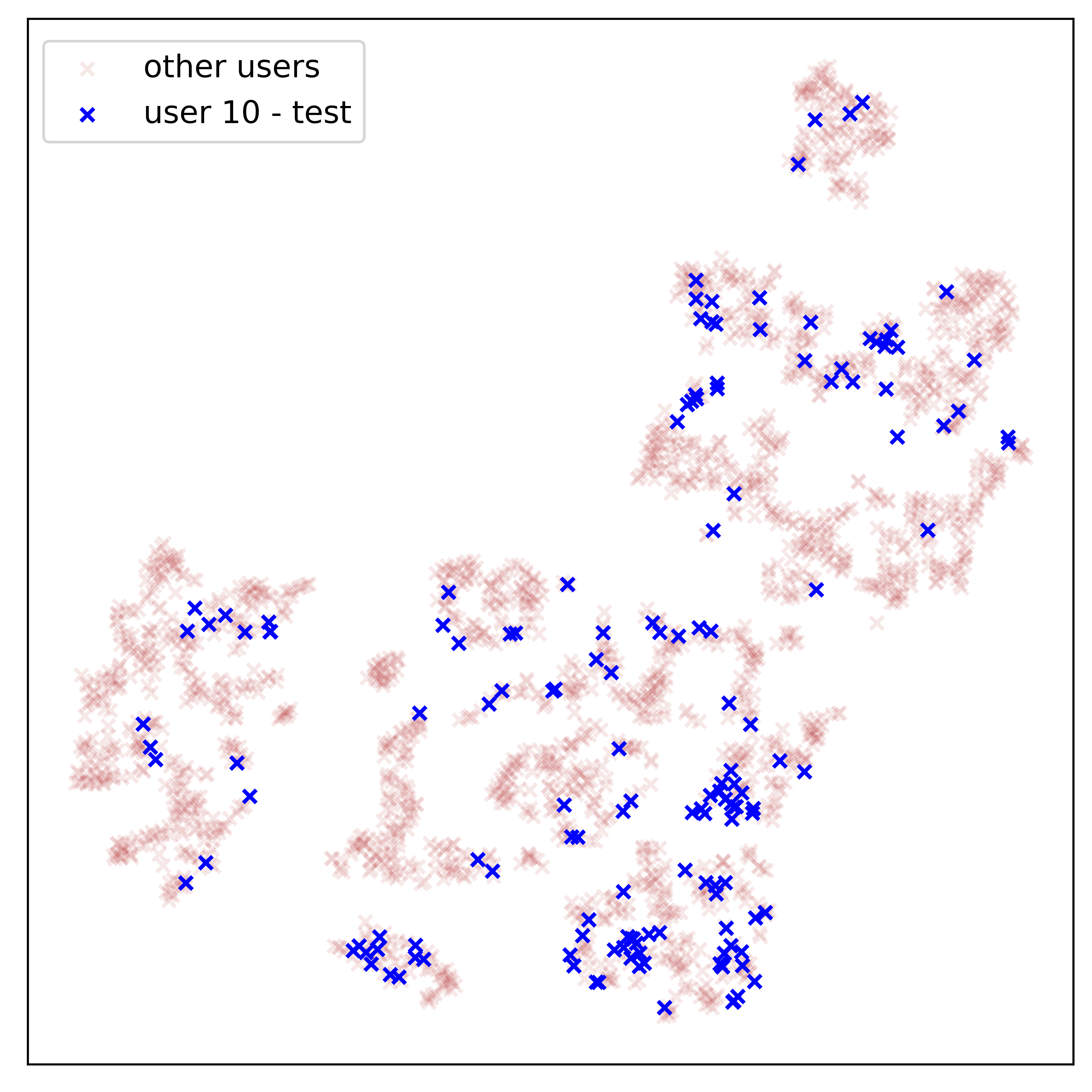}
		\caption[]%
		{{\small User 10}}
	\end{subfigure}
	\hfill
	\caption[]%
	{{\small t-SNE visualisation of latent space embeddings of test data for user $ x $ and training data for other users, for a VAE trained without user $ x $ ($ x = 7 $ or $ x = 10$).}}
	\label{fig:usertsne}
\end{figure*}

We visualise the embeddings of test gestures for users 7 and 10 in Figure \ref{fig:usertsne}. In the case of user 7, our VAE model trained without user 7's gestures learnt a user clustering that is highly relevant to user 7. For user 10, no such effective organisation of the latent space took place. This example shows the importance of the assumption that close together points are more likely to share a user. It is a positive finding that when this assumption is met, synthetic gesture data is of high quality and improves classification performance.

\pagebreak
\pagebreak
\subsection{Reducing enrolment burden} \label{sec:6reduceburden}

\begin{figure*}[t] 
	\centering
	\begin{subfigure}[b]{0.45\textwidth}
		\centering
		\includegraphics[width=\textwidth]{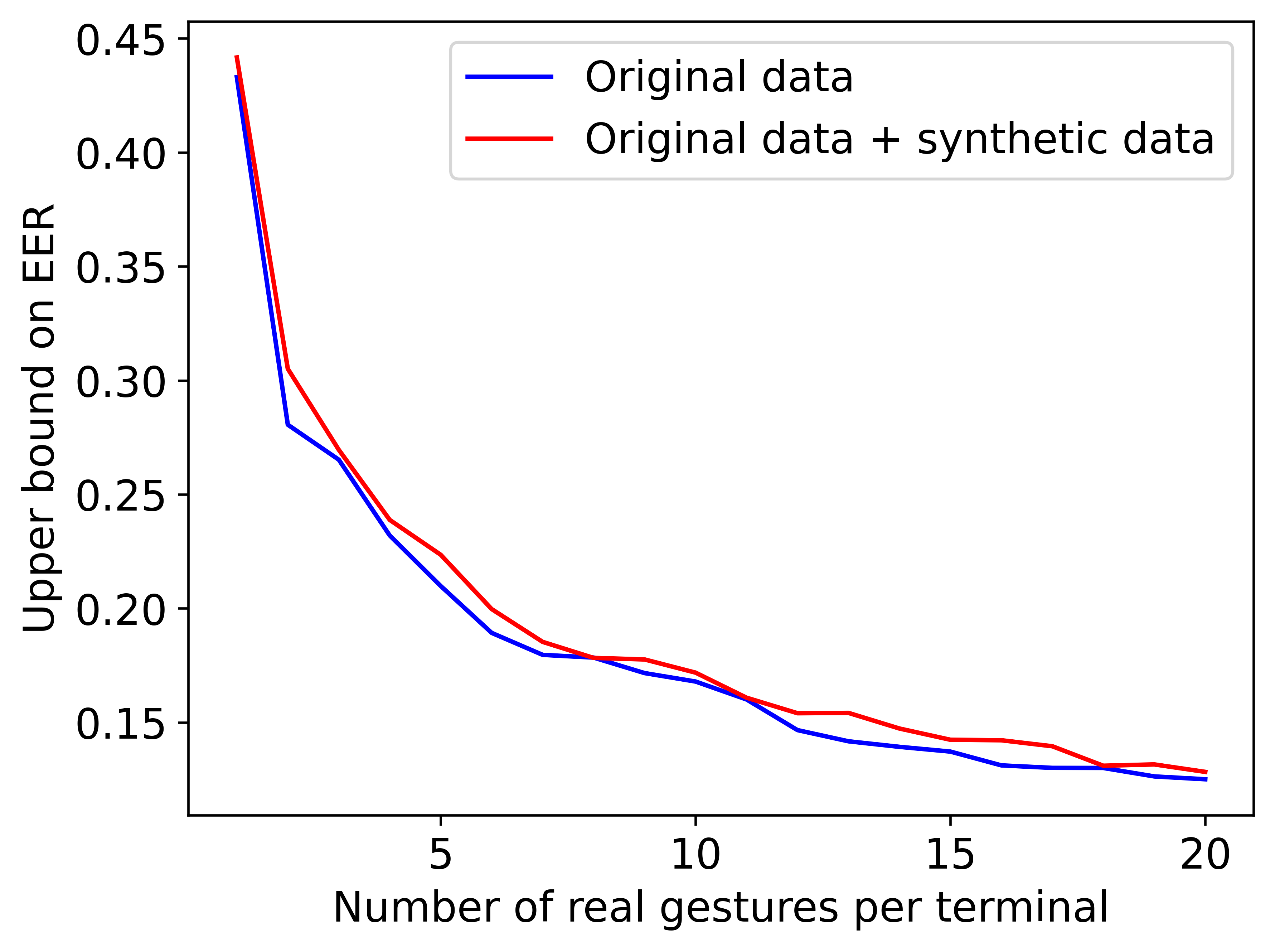}
		\caption[]%
		{{\small Upper bound on EER}}
	\end{subfigure}
	\hfill
	\begin{subfigure}[b]{0.45\textwidth}
		\centering 
		\includegraphics[width=\textwidth]{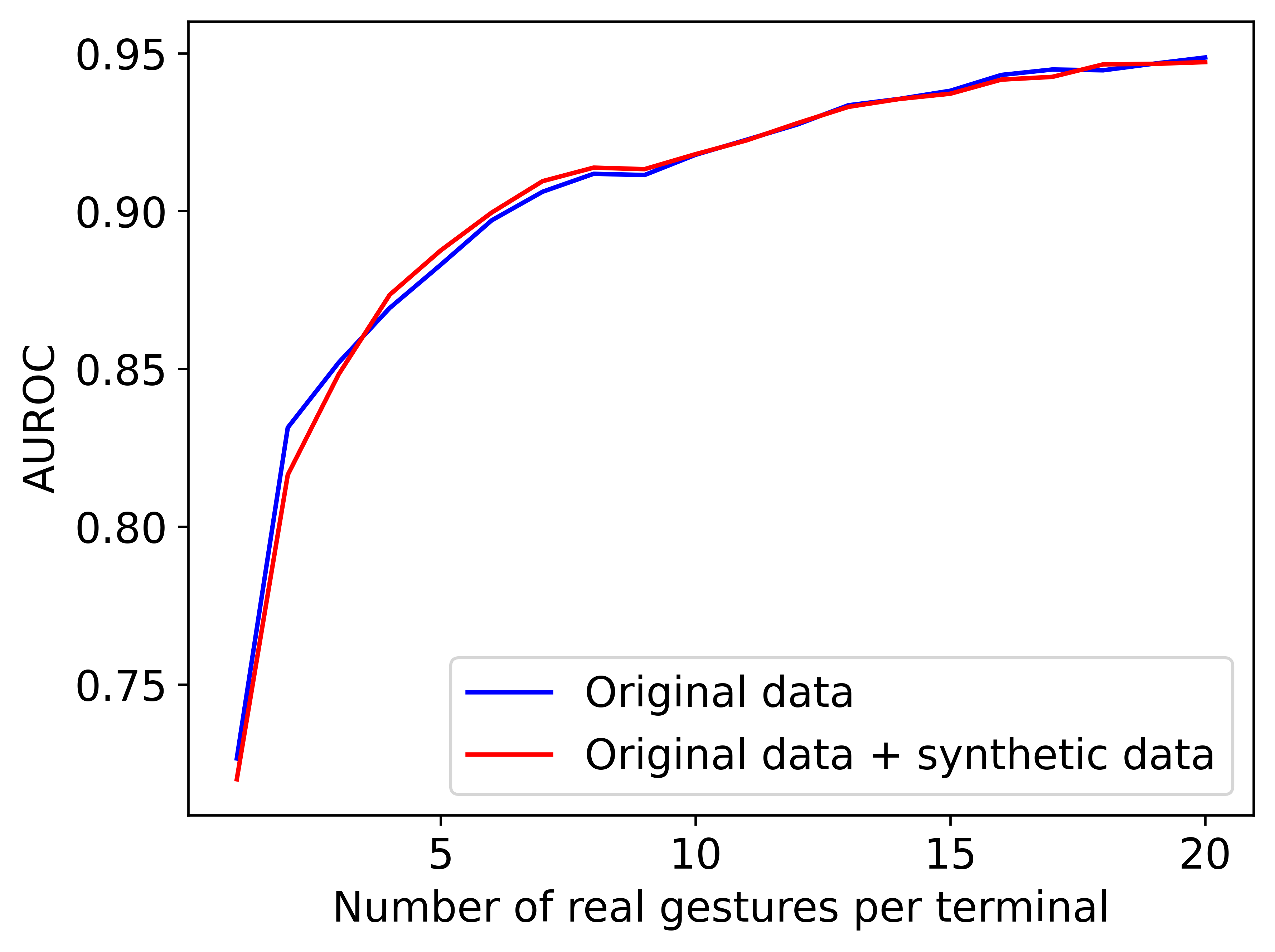}
		\caption[]%
		{{\small AUROC}}
	\end{subfigure}
	\hfill
	\begin{subfigure}[b]{0.7\textwidth}
		\centering 
		\includegraphics[width=\textwidth]{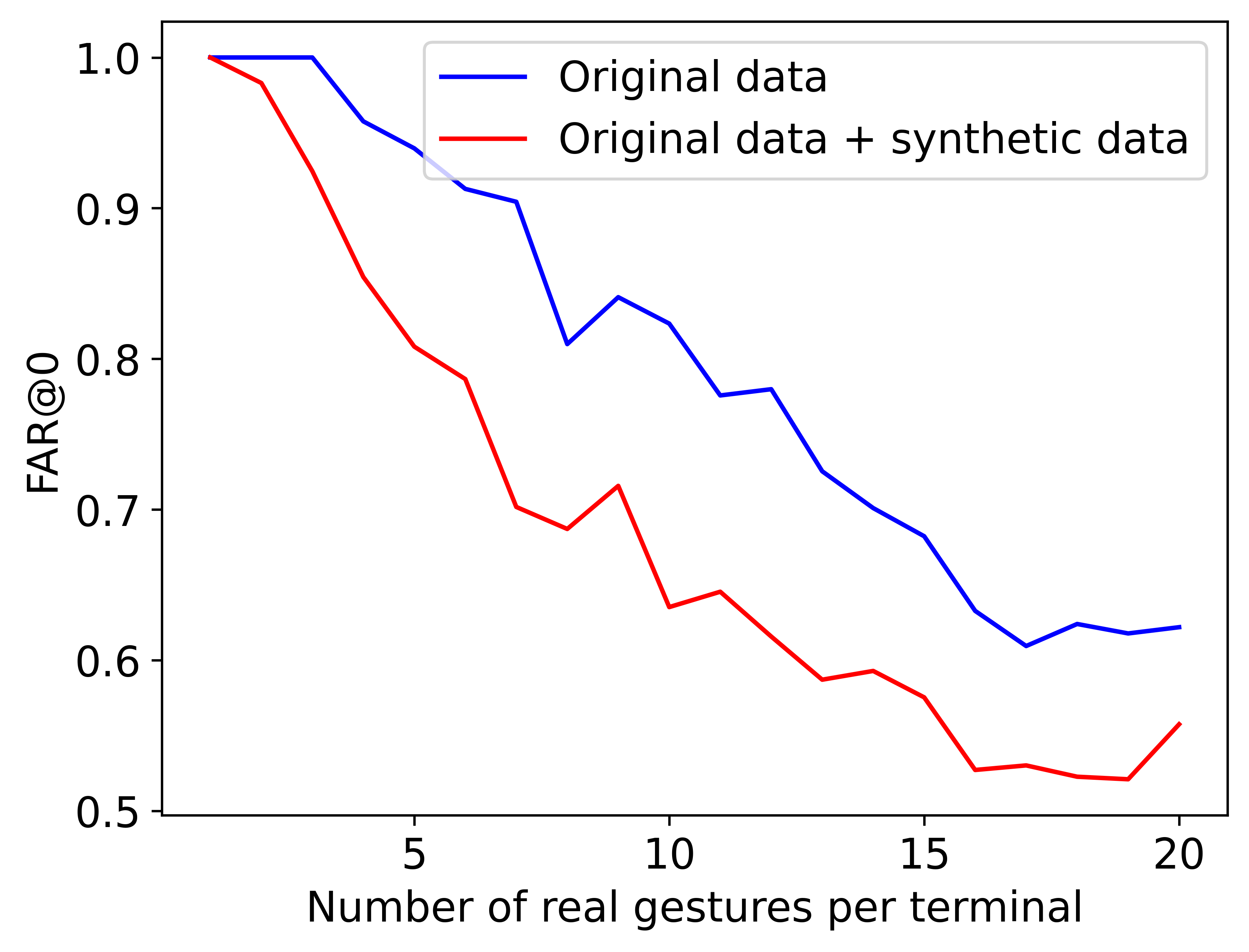}
		\caption[]%
		{{\small FAR@0}}
	\end{subfigure}
	\hfill
	\caption[]%
	{{\small Authentication metrics for random forest classifiers given increasing quantities of training data, with and without the addition of synthetic data.}}
	\label{fig:reduceburden}
\end{figure*}

The primary aim of this work was to explore whether deep learning can reduce the enrolment burden of WatchAuth. To this end, in Figure \ref{fig:reduceburden} we visualise how increasing the number of real user gestures used in training affects the quality of an authentication classifier.

We observe that the addition of synthetic gestures does not adversely affect the EER and AUROC of our trained classifier. Furthermore, the FAR@0 is markedly improved.

It is clear from Figure \ref{fig:reduceburden} that the enrolment burden on a user can be reduced with our methodology. For example, with synthetic gestures, only 9 gestures per terminal were required to reach a FAR@0 value of 70\%, while without synthetic gestures 16 gestures per terminal were needed, an improvement of 42 fewer gestures (in total).

\section{Discussion}

The success of our synthetic data generation methodology rests strongly on the assumption that we can find a latent space of gesture data embeddings in which the gestures of unseen users cluster meaningfully. This work has not fully validated that assumption. However, the success of synthetic gesture generation for some users suggests that with a larger pool of users we would be able to find such a latent space.

Synthetic gestures we generate will inherently share characteristics with gestures from the training data. This is potentially problematic for a security application. However, this is not a critical issue for our particular use case, because WatchAuth is a secondary security measure (the primary measure being the presence of the smartwatch on a user's wrist). Any security WatchAuth can add without compromising usability is beneficial, even if training with synthetic gestures is theoretically likely to allow systematic fraudulent access for the few users who have highly similar gestures to the target user. If present, such a problem would not persist indefinitely, because the authentication system can be retrained once the user has made sufficiently many real gestures.

Both in this chapter and in Chapter \ref{ch:4-auth_improvements}, we observed large differences in the quality of authentication for different users. This trend was common to all models, including WatchAuth. This could be due to fundamental differences between the distinctiveness and stability of user gestures, because some users are more similar to others in the dataset, or a result of improper data collection.



\chapter{Conclusion}\label{ch:7-conclusion}

\minitoc

\section{Conclusions}

In this dissertation, we firstly showed the applicability of deep learning methods to the WatchAuth dataset by designing models that outperform the state-of-the-art for user authentication. We then constructed and iteratively improved a deep generative modelling solution, which we showed could recreate realistic payment gestures for a user. Finally, we demonstrated that synthetic data generated using our approach can significantly reduce the number of gestures required for a user to enrol into a machine learning authentication system.

While this work has successfully proved the concept of augmenting WatchAuth's training process with synthetic data, it should be noted that the level of improvement yielded by adding synthetic data varied strongly between users.

\section{Future work}

\subsection{Larger studies}

Several assumptions made in this work are not fully validated at present; repeating this work with additional data could identify if the assumptions currently fail due to a small study size or fundamentally do not hold. Conducting a study large enough to fully exploit the potential of deep learning would likely require cooperation from industry partners.

\subsection{Deep learning architecture improvements}

To investigate whether Soft-DTW can be a viable reconstruction loss, it is natural to try alternative formulations of Soft-DTW that penalise both the amplitude difference between warped sequences and the level of warping itself to discourage noisy reconstructions.

More fundamentally, the reconstruction losses considered in Section \ref{sec:5losses} cannot directly capture semantic differences between gestures (for example, relative levels of jittering at different timesteps). With additional data, conditional GANs could be employed instead of autoencoder models. To enable user-specific gestures to be generated, the GAN generator would be conditioned on a user's input gesture. The GAN discriminator could then be trained to take in two gestures and classify whether or not they belong to the same user.

Alternatively, a specialised architecture designed for zero-shot learning following Gao et al. could be employed \cite{gaoZeroVAEGANGeneratingUnseen2020}.



\subsection{Teaching users}

Considering the intent-to-pay scenario, we could use deep learning to identify the "best" gesture for signalling intent to pay. We do this by inverting the standard optimisation problem: we take a trained model with fixed weights, and optimise the likelihood of acceptance over possible input gestures. This could have the application of teaching users to make more identifiable payment gestures that minimise false rejections.

\subsection{Extending threat model}

The present threat model assumes a zero-effort attack. It is a natural extension to consider a more sophisticated adversary that attempts to impersonate a user's gesture to make a payment, and to investigate the effectiveness of the methods discussed in this more challenging setting.

\section{Personal development}

While I had prior experience with deep learning, the cybersecurity domain was entirely unfamiliar to me. In particular, it took some adjustment to appreciate the difficulty of fairly evaluating authentication models. Several evaluation metrics were used (e.g. accuracy, F-score) and discarded before selecting the metrics presented in this report.

In this project, both the problem formulation and approach were novel. This made for a challenging research project with little directly applicable literature to guide progress. The narrative presented in this report does not reflect the numerous dead ends explored in Michaelmas term before finding successful deep learning architectures, including difficulty finding hyperparameters and model architectures that would train effectively with such limited data. In reality, I first explored what was possible with the data available during Michaelmas term before re-running experiments and making the series of evidence-based decisions presented in this dissertation. Some of this process could have been avoided by giving myself a more thorough grounding in the literature prior to starting experimenting; some of this was necessary experience learning how to troubleshoot training deep learning models.

\startappendices

\chapter{\label{app:1-model diagrams}Model architectures}


\begin{figure}
	\centering
	\includegraphics[height=0.9\textheight]{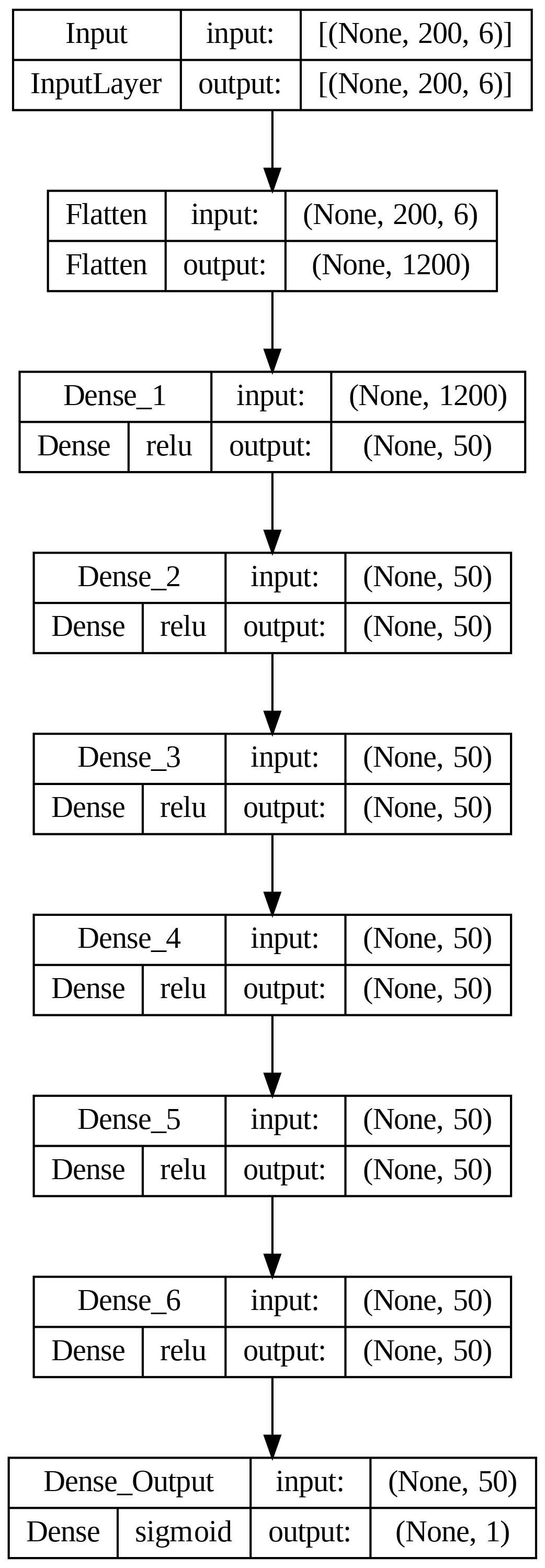}
	\caption{\textit{MLP}}
	\label{fig:mlparchitecture}
\end{figure}

\begin{figure}
	\centering
	\includegraphics[height=0.9\textheight]{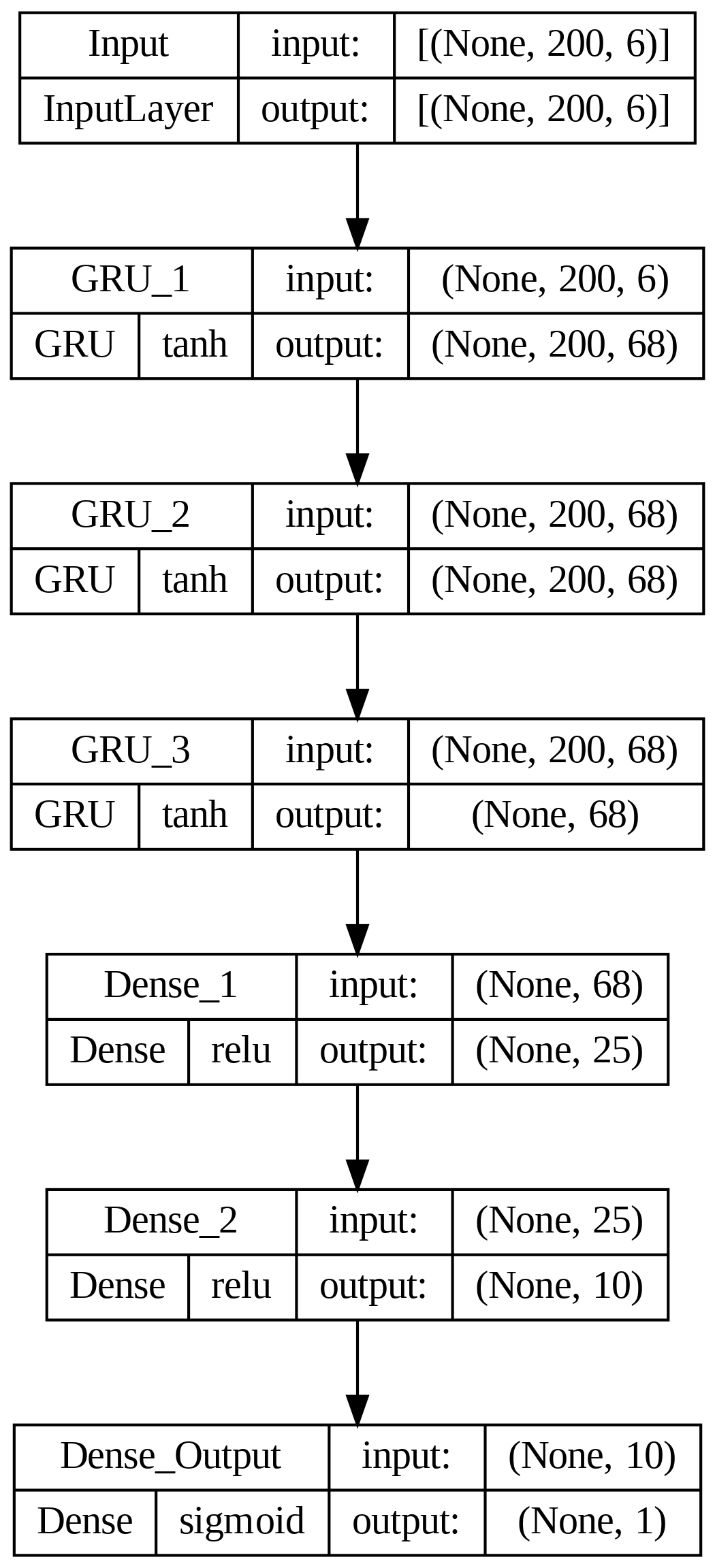}
	\caption{\textit{GRU}}
	\label{fig:gruarchitecture}
\end{figure}

\begin{figure}
	\centering
	\includegraphics[height=0.9\textheight]{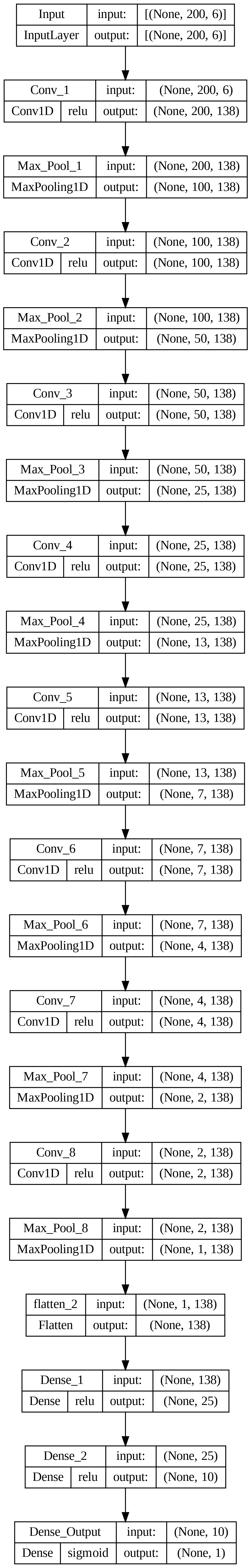}
	\caption{\textit{Conv}}
	\label{fig:convarchitecture}
\end{figure}

\begin{figure}
	\centering
	\includegraphics[height=0.9\textheight]{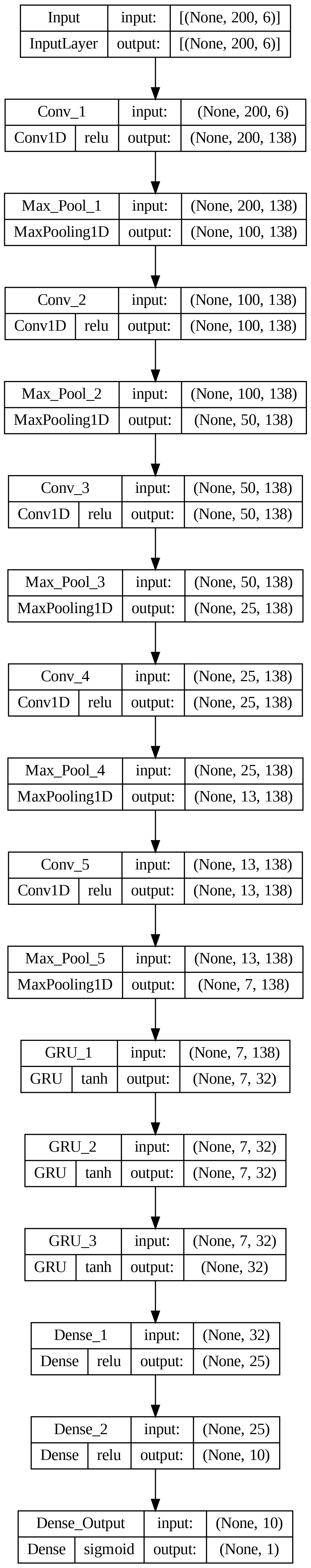}
	\caption{\textit{SimpleMix}}
	\label{fig:simplemixarchitecture}
\end{figure}

\begin{figure}
	\centering
	\includegraphics[height=0.9\textheight]{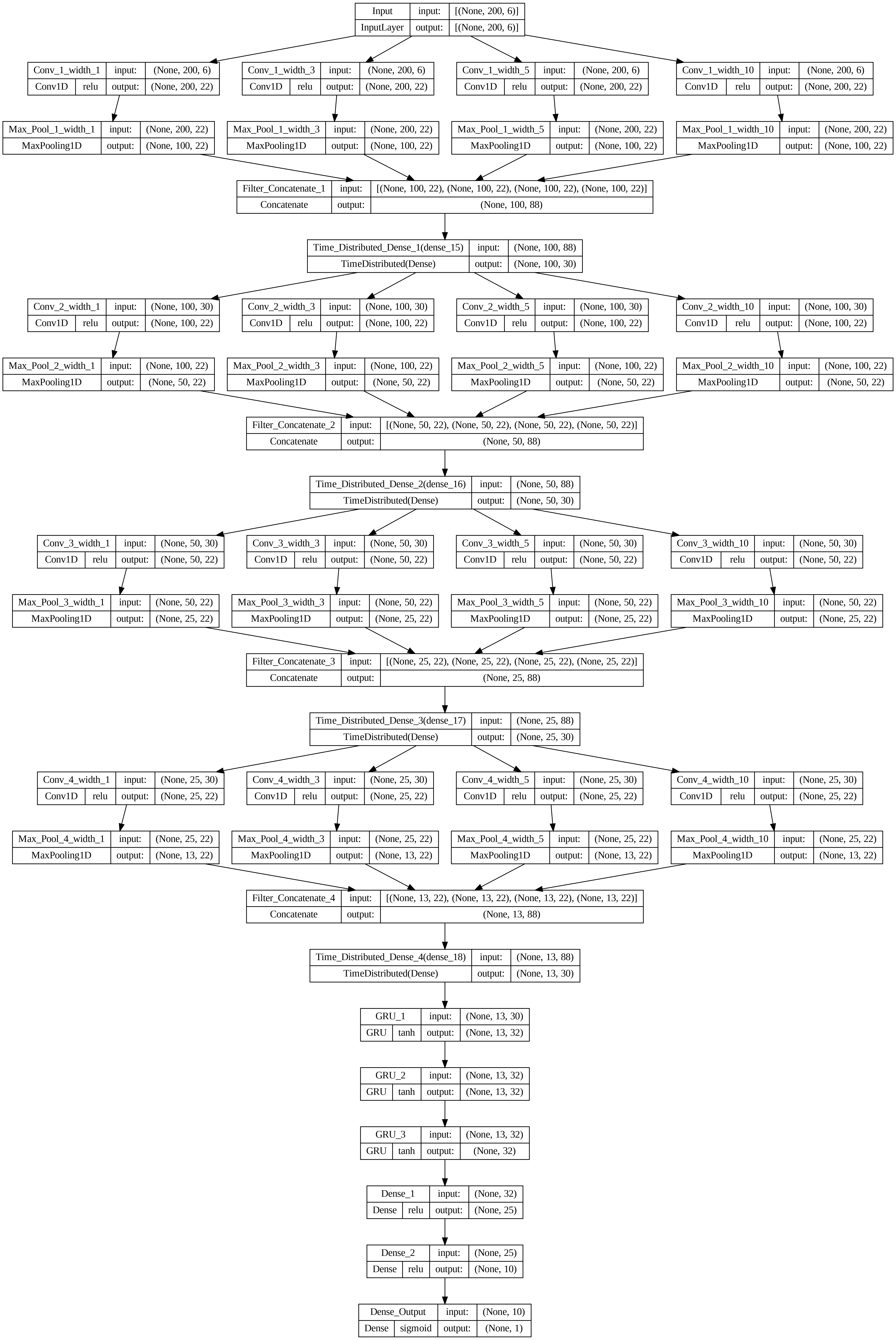}
	\caption{\textit{ComplexMix}}
	\label{fig:complexmixarchitecture}
\end{figure}

\setlength{\baselineskip}{0pt}

{\renewcommand*\MakeUppercase[1]{#1}%
\printbibliography[heading=bibintoc,title={\bibtitle}]}

\end{document}